\documentclass[a4paper,11pt]{article}
\pdfoutput=1
\usepackage{jheppub}

\usepackage{slashed}
\usepackage{bm,wasysym}
\usepackage[normalem]{ulem}

\def\slash#1{\setbox0=\hbox{$#1$}\dimen0=\wd0
      \setbox1=\hbox{/} \dimen1=\wd1 \ifdim\dimen0>\dimen1
      \rlap{\hbox to \dimen0{\hfil/\hfil}} #1                        \else
      \rlap{\hbox to \dimen1{\hfil$#1$\hfil}}
      /   \fi}

\newcommand{\lsim}{
\mathrel{\hbox{\rlap{\hbox{\lower4pt\hbox{$\sim$}}}\hbox{$<$}}}}

\newcommand{\gsim}{
\mathrel{\hbox{\rlap{\hbox{\lower4pt\hbox{$\sim$}}}\hbox{$>$}}}}

\allowdisplaybreaks[2]

\title{Light-Cone Distribution Amplitudes \\ for Heavy-Quark Hadrons}

\author[a]{Guido Bell,}
\author[b]{Thorsten Feldmann,}
\author[c]{Yu-Ming Wang,}
\author[d]{and Matthew W~Y~Yip}

\affiliation[a]{Rudolf Peierls Centre for Theoretical Physics, University of Oxford,
Oxford,
UK}

\affiliation[b]{Theoretische Elementarteilchenphysik,
Universit\"at Siegen, Siegen, Germany}

\affiliation[c]{{
Institut f\"ur Theoretische Physik~E, RWTH Aachen, Aachen, Germany; \newline
Physik Department T31,
Technische Universit\"at M\"unchen, Garching, Germany}}

\affiliation[d]{Institute for Particle Physics Phenomenology,
Durham University, Durham, UK}

\emailAdd{guido.bell@physics.ox.ac.uk}
\emailAdd{thorsten.feldmann@uni-siegen.de}
\emailAdd{yuming.wang@tum.de}
\emailAdd{m.w.yip@durham.ac.uk}

\abstract{We construct parametrizations of light-cone distribution
amplitudes (LCDAs) for $B$-mesons and $\Lambda_b$-baryons that obey various theoretical 
constraints, and which are simple to use in factorization theorems
relevant for phenomenological applications in heavy-flavour physics.
In particular, we find the eigenfunctions of the Lange-Neubert renormalization kernel,
which allow for a systematic implementation of renormalization-group evolution effects
for both $B$-meson and $\Lambda_b$-baryon decays.
We also present a new strategy to construct LCDA models from momentum-space projectors,
which can be used to implement Wandzura-Wilczek--like
relations, and which allow for a comparison with
theoretical approaches that go beyond the collinear limit for
the light-quark momenta in energetic heavy-hadron
decays.
}

\keywords{Heavy Quarks, Light-Cone Distribution Amplitudes, Renormalization}

\note{\today.\newline
Oxford: OUTP-13-15P; Siegen: SI-HEP-2013-05, QFET-2013-05; Aachen: TTK-13-19, \newline SFB/CPP-13-60;
Munich: TUM-HEP-899/13; Durham: IPPP/13/55, DCPT/13/110.}

\begin{document}

\maketitle


\section{Motivation}

Light-cone distribution amplitudes (LCDAs) play a key role in the factorization of short-
and long-distance dynamics entering
\emph{exclusive} transition amplitudes in Quantum Chromodynamics (QCD). Defined as hadronic matrix elements
of composite QCD operators with fields separated along the light-cone, they represent
genuine non-perturbative quantities, whose renormalization-scale dependence follows from
the anomalous dimension of the defining operators which can be computed in perturbation
theory. While early analyses focused on light hadrons (pions, kaons, nucleons, etc.\
\cite{Efremov:1979qk,Lepage:1979zb,Duncan:1979hi,Chernyak:1983ej}), with the running $b$-physics program
at ``{\it B}-Factories'' or high-luminosity
hadron colliders, the interest in exclusive decay channels of $B$-mesons and $\Lambda_b$-baryons
revealed the importance of LCDAs for hadrons containing a heavy quark.
For instance, the former provide the dominant hadronic input for
the theoretical description of radiative leptonic $B$-decays
\cite{Korchemsky:1999qb,DescotesGenon:2002mw,Lunghi:2002ju,Bosch:2003fc,Beneke:2011nf,Braun:2012kp},
they enter the QCD-factorization approach
for non-leptonic and semi-leptonic $B$-decays
\cite{Beneke:1999br,Beneke:2000wa,Bosch:2001gv,Ali:2001ez,Beneke:2001at,Kagan:2001zk},
and they naturally appear in the context of soft-collinear effective theory
(SCET) \cite{Bauer:2000yr,Beneke:2002ph}. They can also be used in
correlation functions that form the basis of light-cone sum rules for heavy-to-light
transition form factors
\cite{DeFazio:2005dx,Khodjamirian:2005ea}, and they appear as limiting cases in the $k_T$-factorization approach for
heavy-meson \cite{Keum:2000wi,Lu:2000em} or heavy-baryon \cite{He:2006ud} decays.

Theoretical properties of LCDAs
have been classified for heavy mesons (see e.g.\ \cite{Grozin:1996pq,Beneke:2000wa,Kawamura:2001jm})
and heavy baryons (see e.g.\ \cite{Ball:2008fw,Feldmann:2011xf,Ali:2012pn}).
The renormaliza\-tion-group (RG) evolution of the LCDAs has been derived in the heavy-quark limit,
where the $b$-quark fields are treated in heavy-quark effective theory (HQET),
for both, mesons \cite{Lange:2003ff,Braun:2003wx,Lange:2004yh,Lee:2005gza,Bell:2008er,DescotesGenon:2009hk,Knodlseder:2011gc}
and baryons \cite{Ball:2008fw}.
Available model parametrizations are typically inspired by sum-rule analyses, for instance
\cite{Braun:2003wx} for $B$-mesons or \cite{Ball:2008fw} for $\Lambda_b$-baryons.
Certain inverse moments of the $B$-meson LCDA can also be constrained from experimental data, most notably from
the radiative leptonic decay $B\to \gamma\ell\nu$ (see \cite{Beneke:2011nf,Braun:2012kp}
for recent analyses).

In this article, our aim is to improve the theoretical modelling of LCDAs for
$B$-mesons and $\Lambda_b$-baryons in two aspects: First, we introduce a new representation
for  LCDAs in terms of a convolution of Bessel functions and a spectral function,
which
behaves in a dual way compared to the original LCDAs.
This representation diagonalizes the Lange-Neubert (LN) evolution kernel,
and therefore greatly simplifies the analytic solution of
the RG equation. Our formalism also provides a systematic method to include
the Efremov-Radyushkin--Brodsky-Lepage (ERBL) kernel, which arises from gluon
exchange among the light spectator quarks in the $\Lambda_b$-baryon. ---
Second, we develop a general procedure for an efficient modelling of LCDAs,
starting from momentum-space projectors, which satisfy the equations of motion for the
partonic Fock-state components. In this approach, the
LCDAs follow as simple integrals over a set of (fewer) on-shell wave functions,
which allows for a comparison between LCDAs used in the collinear
factorization approach and transverse-momentum-dependent wave-functions
employed in the $k_T$-factorization approach.
For $B$-mesons, our formalism reproduces the so-called
Wandzura-Wilczek (WW) relations established in \cite{Beneke:2000wa}, and for
$\Lambda_b$-baryons, we derive
analogous relations between the various 3-particle LCDAs.

\clearpage 

Our paper is organized as follows.
In the first part, we focus on  $B$-meson LCDAs for which we first
recollect the main definitions and results, before we introduce our
new representation in terms of dual spectral functions and give a detailed discussion
on the RG properties. As a sample application of our formalism, we  
briefly reconsider the RG-improved factorization formula for the 
radiative leptonic $B\to\gamma\ell\nu$ decay.
We proceed with the definition of on-shell momentum-space
wave functions, and show how to construct the LCDAs and the corresponding
momentum-space projectors to be used in applications of (collinear) QCD
factorization. We also discuss how to incorporate corrections to the
WW relations from 3-particle Fock states,
and discuss two simple wave function models in detail. ---
In the second part we present an analogous analysis for baryon
LCDAs. Here, we show that the momentum-space construction -- together with some
simplifying assumptions -- leads to an enormous reduction of independent
hadronic functions. The RG equations for the baryon LCDAs
are more complicated than for the meson LCDAs, because
the LN kernel and the ERBL kernel
depend differently on the two light-quark momenta.
Nevertheless, we show that
the RG equations can be solved systematically in the dual space in an expansion in Gegenbauer polynomials.
This also establishes the general properties of leading-twist
baryon LCDAs  in the  asymptotic limit.

Our results are shortly summarized in Section~\ref{summary}. In the appendix we generalize our
results for the $B$-meson momentum-space projectors to an arbitrary frame, where the heavy quark has a
transverse velocity component.
We also provide some details about the
ERBL evolution of the baryon LCDAs, and collect some general integral relations for Bessel functions.

\section{$B$--Mesons}

\subsection{Light-Cone Distribution Amplitudes}

We first recapitulate the properties of $B$-meson LCDAs
in the heavy-quark limit, see e.g.\
\cite{Grozin:1996pq} and references therein.

\subsubsection{Light-Cone Projector for 2-Particle Fock State}

The 2-particle LCDAs of the $B$-meson in HQET (which differ from the QCD definition,
see \cite{Pilipp:2007sb,Li:2012md})
can be obtained from the coordinate-space matrix elements
\begin{align}
 & \langle 0 | \bar q^\beta(z) \, [z,0] \, h_v^\alpha(0)| \bar B(v)\rangle
 \cr
 &=
  - \frac{i \tilde f_B m_B}{4} \left[ \frac{1+\slashed v}{2} \left\{ 2 \, \tilde \Phi_B^+(t,z^2) +
  \frac{\tilde \Phi_B^-(t,z^2) - \tilde \Phi_B^+(t,z^2)}{t} \, \slashed z \right\} \gamma_5 \right]^{\alpha\beta} \,,
  \label{Bdef}
\end{align}
in the limit of light-like separation $z^2 \to 0$.
Here $t = v\cdot z$, and $[z,0]$ denotes a gauge-link represented by
a straight Wilson line along $z^\mu$. The heavy $b$-quark field in HQET is denoted as $h_v$
for a $B$-meson that moves with velocity $v^\mu$.
The decay constant $\tilde f_B$ in the heavy-quark limit includes the
non-trivial scale dependence from the non-vanishing anomalous dimension of the local heavy-to-light
current in HQET.

The terms in curly brackets can be expanded around $z^2=0$, using the power-counting induced by the
convolution with generic hard-scattering kernels in factorization theorems (see the discussion in \cite{Beneke:2000wa}).
Introducing light-like vectors $n_\pm^\mu$ with $(n_- \!\cdot\! n_+) = 2$ and $v^\mu = \frac{n_-^\mu+n_+^\mu}{2}$,
and taking $(n_-z) \ll z_\perp \ll (n_+z)$,
we obtain\footnote{The generalization to frames where $v^\mu \neq (n_+^\mu+n_-^\mu)/2$ can be found in
Appendix \ref{generalframe}.}
\begin{align}
& 2 \, \tilde \Phi_B^+(t,z^2) +
  \frac{\tilde \Phi_B^-(t,z^2) - \tilde \Phi_B^+(t,z^2)}{t} \, \slashed z
  \cr
\stackrel{\slashed v \to 1}{\longrightarrow} \quad & \tilde \phi_B^+(\tau)\, \slashed n_+ +
  \tilde \phi_B^-(\tau) \, \slashed n_- + \frac{\tilde \phi_B^-(\tau) - \tilde \phi_B^+(\tau)}{\tau} \, \slashed z_\perp
  + {\cal O}(z_\perp^2, n_-z) \,,
  \label{lightB}
\end{align}
where $t \to \tau = \frac{n_+ z}{2}$ can be interpreted as
the Fourier-conjugated variable to the momentum component $\omega=(n_- k)$ associated with the light anti-quark field.
Defining the LCDAs in momentum space from the Fourier transform,
\begin{align}
 \phi_B^\pm(\omega) &\equiv \int \frac{d\tau}{2\pi} \, e^{i \omega \tau} \, \tilde \phi_B^\pm(\tau)  \,,
 \label{phiBfourier}
\end{align}
the light-cone expansion in (\ref{lightB}) corresponds to the momentum-space projector \cite{Beneke:2000wa}
\begin{align}
 {\cal M}_B(v,\omega)   &= - \frac{i \tilde f_B m_B}{4} \left[ \frac{1+\slashed v}{2}
 \bigg\{ \phi_B^+(\omega) \, \slashed n_+ +
  \phi_B^-(\omega) \, \slashed n_-   \right. \cr
 & \qquad\qquad \left. -\int_0^\omega d\eta \left(\phi_B^-(\eta)
   - \phi_B^+(\eta)\right) \gamma^\mu_\perp \frac{\partial}{\partial k_\perp^\mu } \bigg\}   \gamma_5 \right] \,,
  \label{Bproj}
\end{align}
to be used in factorization theorems
with hard-scattering kernels where the $k_\perp^2$ and $(n_+k)$-dependence can be neglected
(see below).\subsubsection{Wandzura-Wilczek Approximation}

In the approximation where 3-particle contributions to the $B$-meson wave function
are neglected, the equations of motion for the (massless) light-quark field
yield WW relations between the 2-particle LCDAs \cite{Beneke:2000wa},
\begin{align}
& \frac{d\tilde \phi_B^-(\tau)}{d\tau} + \frac{1}{\tau} \left( \tilde \phi_B^-(\tau)-\tilde \phi_B^+(\tau)\right) =0 \cr
\Leftrightarrow \quad
&
\int_0^\omega d\eta \left( \phi_B^-(\eta)-\phi_B^+(\eta) \right) = \omega \, \phi_B^-(\omega) \quad
\Leftrightarrow \quad \phi_B^+(\omega) = - \omega \, \frac{d\phi_B^-(\omega)}{d\omega} \,.
 \label{WW}
\end{align}
Including the contributions from 3-particle LCDAs as defined in (\ref{3-particleLCDA}) below,
this is generalized to
\begin{align}
 \omega \, \phi_B^-(\omega) - \int_0^\omega d\eta \left(\phi_B^-(\eta)-\phi_B^+(\eta) \right)
 = 2 \, \int_0^\omega d\eta \int_{\omega-\eta}^\infty \frac{d\xi}{\xi} \, \frac{\partial}{\partial \xi} 
 \left( \Psi_A(\eta,\xi)-\Psi_V(\eta,\xi)\right)
 \,.
 \label{WWcorr}
\end{align}

\subsubsection{Renormalization-Group Evolution}

The RG evolution equation for the LCDA $\phi_B^+(\omega,\mu)$ reads
(with a slight change of notation compared to \cite{Lange:2003ff})
\begin{align}
 \frac{d\phi_B^+(\omega,\mu)}{d\ln\mu}
 &= - \left[\Gamma_{\rm cusp}(\alpha_s) \, \ln \frac{\mu}{\omega}
  + \gamma_+(\alpha_s) \right] \phi_B^+(\omega,\mu)  - \omega \, \int_0^\infty d\eta \, 
  \Gamma_{+}(\omega,\eta,\alpha_s) \, \phi_B^+(\eta,\mu)\,,
\end{align}
where the leading terms in the various contributions to the anomalous dimensions
in units of $\alpha_s C_F/4\pi$ are
\begin{align}
& \Gamma_{\rm cusp}^{(1)} =4 \,, \qquad
\gamma_+^{(1)} = -2 \,, \qquad
\Gamma^{(1)}_{+}(\omega,\eta) = - \Gamma_{\rm cusp}^{(1)} \left[ \frac{\theta(\eta-\omega)}{\eta(\eta-\omega)}
+ \frac{\theta(\omega-\eta)}{\omega(\omega-\eta)} \right]_+ \,,
\label{one-loop}
\end{align}
with the usual definition of the plus distribution \cite{Lange:2003ff}.
The RG equation can be solved in
closed form \cite{Lee:2005gza}.\footnote{The evolution kernel has been calculated at one-loop,
and in \cite{Lee:2005gza} it has been conjectured that the structure of the evolution kernel
remains a general feature at higher orders in the perturbative analysis.}
Starting from the Fourier transform\footnote{This can also be viewed as Mellin moments
$\langle \omega^{N-1}\rangle_B^+$ for $N=-i\theta$.}
with respect to the variable $\ln \omega/\mu$,
\begin{align}
 \varphi_B^+(\theta,\mu) &=  \int_0^\infty \frac{d\omega}{\omega} \,\left( \frac{\omega}{\mu} \right)^{-i\theta} \phi_B^+(\omega,\mu)
  \quad \Leftrightarrow \quad
 \phi_B^+(\omega,\mu)  = \int_{-\infty}^\infty \frac{d\theta}{2\pi} \left(\frac{\omega}{\mu} \right)^{i\theta} \varphi_B^+(\theta,\mu)
 \,,
\end{align}
one has an explicit solution of the RG equation,
\begin{align}
 \varphi_B^+(\theta,\mu) &= e^{V-2\gamma_Eg} \left(\frac{\mu}{\mu_0} \right)^{i\theta}
  \, \frac{\Gamma(1-i\theta) \, \Gamma(1+i\theta-g)}{\Gamma(1+i\theta)\, \Gamma(1-i\theta +g)}
  \, \varphi_B^+(\theta +ig,\mu_0) \,,
\end{align}
where the RG functions are \cite{Lee:2005gza}
\begin{align}
 V:=V(\mu,\mu_0) & = - \int\limits_{\alpha_s(\mu_0)}^{\alpha_s(\mu)} \frac{d\alpha}{\beta(\alpha)}
 \left[\Gamma_{\rm cusp}(\alpha) \int\limits_{\alpha_s(\mu_0)}^\alpha \frac{d\alpha'}{\beta(\alpha')}
 + \gamma_+(\alpha) \right] \,,\nonumber
 \\
 g := g(\mu,\mu_0) &=  \int\limits_{\alpha_s(\mu_0)}^{\alpha_s(\mu)} d\alpha \,
 \frac{\Gamma_{\rm cusp}(\alpha)}{\beta(\alpha)} \,.
 \label{RGfunctions}
\end{align}
Transforming back to momentum space, the RG solution can be written as
a convolution involving hypergeometric functions,\footnote{The corresponding
formalism for the LCDA $\tilde \phi_B^+(\tau,\mu)$ in coordinate space
has been worked out in \cite{Kawamura:2010tj}.}
\begin{align}
 \phi_B^+(\omega,\mu)& = e^{V-2\gamma_E g} \, \frac{\Gamma(2-g)}{\Gamma(g)}
 \, \int_0^\infty \frac{d\eta}{\eta} \, \phi_B^+(\eta,\mu_0)
 \left( \frac{\max(\omega,\eta)}{\mu_0} \right)^g \cr & \qquad {} \times
 \frac{\min(\omega,\eta)}{\max(\omega,\eta)} \,
 {}_2F_1\left(1-g,2-g,2,\frac{\min(\omega,\eta)}{\max(\omega,\eta)}\right)\,,
\label{LeeNeubertsol}
\end{align}
which is valid for $0<g<1$ (larger values of $g$ do not
appear in phenomenological applications and will not be considered further).

As one of the central new ideas of this paper,
we suggest an alternative representation of the RG solution, which is
obtained from the ansatz
\begin{align}
 \varphi_B^+(\theta,\mu) &:= \frac{\Gamma(1-i\theta)}{\Gamma(1+i\theta)} \,
 \int_0^\infty \frac{d\omega'}{\omega'} \, \rho_B^+(\omega',\mu)
 \left(\frac{\mu}{\omega'} \right)^{i\theta} \,.
\end{align}
The particular parametrization for $\varphi_B^+(\theta)$ implies a simple RG behaviour
for the spectral function $\rho_B^+(\omega')$,
\begin{align}
 \rho_B^+(\omega',\mu) &=  e^{V} \left(\frac{\mu_0}{\hat \omega'} \right)^{-g} \, \rho_B^+(\omega',\mu_0)
= e^{\bar V} \left(\frac{\mu \mu_0}{(\hat \omega')^2} \right)^{-g/2} \, \rho_B^+(\omega',\mu_0)\,,
 \label{BmesonRGE1}
\end{align}
as a solution to the RG equation,
\begin{align}
 \frac{d\rho_B^+(\omega',\mu)}{d\ln\mu}
 &= - \left[\Gamma_{\rm cusp}(\alpha_s) \, \ln \frac{\mu}{\hat\omega'} + \gamma_+(\alpha_s) \right]
 \rho_B^+(\omega',\mu) \,.
\end{align}
Here we have defined
\begin{align}
&
 \hat \omega' = e^{-2\gamma_E} \, \omega' \,,
 \qquad \bar V(\mu,\mu_0) = \frac12 \left( V(\mu,\mu_0)-V(\mu_0,\mu) \right)\,,
 \label{hatVdef}
\end{align}
to write the solution in a manifestly symmetric form with respect to $\mu \leftrightarrow \mu_0$.
The relation between the original LCDA $\phi_B^+(\omega)$ and the spectral function  $\rho_B^+(\omega')$ then follows
as\footnote{In the following,
we implicitly assume that $\rho_B^+(\omega',\mu_0) \sim 1/(\omega')^\epsilon$ with $\epsilon>0$ in the limit $\omega'\to\infty$,
such that the integrals in (\ref{phiBrhoB}) converge for $g-\epsilon<1$.}
\begin{align}
 \phi_B^+(\omega,\mu) &= \int_{-\infty}^\infty \frac{d\theta}{2\pi}
 \frac{\Gamma(1-i\theta)}{\Gamma(1+i\theta)} \, \int_0^\infty \frac{d\omega'}{\omega'} \, \rho_B^+(\omega',\mu) \left(\frac{\omega}{\omega'} \right)^{i\theta}
 \cr &=   \int_0^\infty \frac{d\omega'}{\omega'} \, \sqrt{\frac{\omega}{\omega'}} \, J_1\left( 2 \, \sqrt{\frac{\omega}{\omega'}}\right)  \rho_B^+(\omega',\mu) \cr
 &=  e^{V} \,  \int_0^\infty \frac{d\omega'}{\omega'} \, \sqrt{\frac{\omega}{\omega'}} \,
 J_1\left( 2 \, \sqrt{\frac{\omega}{\omega'}}\right) \left(\frac{\mu_0}{\hat\omega'}
 \right)^{-g} \, \rho_B^+(\omega',\mu_0)
 \,,
 \label{phiBrhoB}
\end{align}
where $J_1$ is the Bessel function of the first kind.
The expansion of the LCDA $\phi_B^+(\omega,\mu)$ in terms of Bessel functions and a
spectral weight $\rho_B^+(\omega',\mu)$ with simple RG properties can be viewed as the
analogue of the Gegenbauer expansion for light mesons, where the Gegenbauer polynomials
diagonalize the corresponding one-loop RG kernel. 
Therefore, instead of defining a model for the input function $\phi_B^+(\omega,\mu_0)$, one
can equivalently define a model for the dual spectral function $\rho_B^+(\omega',\mu_0)$
at a given hadronic input scale, and determine $\phi_B(\omega,\mu)$ at a different scale
from a relatively simple convolution integral.
Alternatively, one can express $\rho_B^+(\omega',\mu_0)$ in terms of $\phi_B^+(\omega,\mu_0)$ as
\begin{align}
 \rho_B^+(\omega',\mu_0) &=
 \int_0^\infty \frac{d\omega}{\omega} \, \sqrt{\frac{\omega}{\omega'}} \,
 J_1\left( 2 \, \sqrt{\frac{\omega}{\omega'} }\right) \phi_B^+(\omega,\mu_0) \,,
 \label{rhoBexpr}
\end{align}
such that the solution of the RG equation for $\phi_B^+(\omega,\mu)$ can also be obtained
from a double convolution,
\begin{align}
 \phi_B^+(\omega,\mu) &=e^{V} \,
 \int_0^\infty \frac{d\omega'}{\omega'} \,
 \int_0^\infty \frac{d\eta}{\eta} \,
 \sqrt{\frac{\omega \eta}{(\omega')^2}} \,
 \cr & \qquad {}\times
  J_1\left( 2 \, \sqrt{\frac{\eta}{\omega'} }\right)
  J_1\left( 2 \, \sqrt{\frac{\omega}{\omega'}}\right) \left(\frac{\mu_0}{\hat\omega'}
 \right)^{-g} \, \phi_B^+(\eta,\mu_0) \,.
 \label{phiBRGsol}
\end{align}

It is also interesting to note, how the spectral function is connected with the
function $\tilde \phi_B^+(\tau)$ appearing in the defining light-cone matrix
elements. From (\ref{phiBfourier}) and (\ref{rhoBexpr}) we obtain
\begin{align}
 \rho_B^+(\omega',\mu) &=  \int \frac{d\tau}{2\pi} \left( 1 - \exp\left[-\frac{i}{ \omega'\,\tau} \right]
 \right) \, \tilde \phi_B^+(\tau,\mu) \,.
\end{align}
The fact that the product $\omega'\tau$ appears as the \emph{inverse} in the exponential
--- as compared to $\omega\tau$ in the conventional Fourier transform in
(\ref{phiBfourier}) --- justifies the notion ``dual function''.
It also illustrates, why the function $\rho_B^+(\omega')$ cannot be reconstructed
through its positive moments related
to a local operator product expansion
around $\tau \to 0$ (see also the discussion in \cite{Braun:2003wx,Lee:2005gza,Kawamura:2008vq}).

The 2-particle operator that defines the
LCDA $\phi_B^+(\omega,\mu)$ does not mix with the contributions from
3-particle operators under RG evolution \cite{DescotesGenon:2009hk,Knodlseder:2011gc}.
In contrast, the evolution of the LCDA $\phi_B^-(\omega,\mu)$ contains
a part that is independent of the 3-particle contributions and
can be reconstructed from the WW relation \cite{Bell:2008er},
and a part that describes the explicit mixing with the combination
$(\Psi_A - \Psi_V)$ of
3-particle LCDAs  that enters the corrections
to the WW relations in (\ref{WWcorr}).
In the WW approximation, the RG  equation for the LCDA $\phi_B^-(\omega,\mu)$
can be solved in closed form \cite{Bell:2008er}. In terms of our new strategy for the RG solution,
we find that the transformation
\begin{align}
 \phi_B^-(\omega,\mu) &=
 \int_0^\infty \frac{d\omega'}{\omega'} \, J_0\left( 2 \, \sqrt{\frac{\omega}{\omega'}}\right)  \rho_B^-(\omega',\mu)
 \label{phiBrhoBm}
\end{align}
with inverse transformation
\begin{align}
 \rho_B^-(\omega',\mu) &=
 \int_0^\infty \frac{d\omega}{\omega'} \,
 J_0\left( 2 \, \sqrt{\frac{\omega}{\omega'} }\right) \phi_B^-(\omega,\mu)
\end{align}
diagonalizes the corresponding one-loop renormalization kernel.
The WW relation (\ref{WW})  implies a particularly simple relation
between the spectral functions,
\begin{align}
 \rho_B^-(\omega') &= \rho_B^+(\omega'),
 \end{align}
 and the solution for the LCDA $\phi_B^-(\omega,\mu)$ takes the form
\begin{align}
 \phi_B^-(\omega,\mu)  &=  e^{V} \,  \int_0^\infty \frac{d\omega'}{\omega'} \,
 J_0\left( 2 \, \sqrt{\frac{\omega}{\omega'}}\right) \left(\frac{\mu_0}{\hat\omega'}
 \right)^{-g} \, \rho_B^-(\omega',\mu_0)
 \,.
\end{align}
We will give explicit examples for $\phi_B^\pm(\omega)$, and $\rho_B^\pm(\omega')$
from different models further below.

\subsubsection{Application in Factorization Theorems}

The most relevant parameter in phenomenological applications
of the factorization approach is the first inverse moment of the LCDA $\phi_B^+(\omega)$.
Interestingly, equation (\ref{phiBRGsol}) implies ---
as a consequence of the completeness relation (\ref{complete}) for Fourier-Bessel transforms --- 
that this moment is identical to the first inverse moment of the corresponding spectral function $\rho_B^+(\omega')$.
This result can be generalized to moments with one or two additional
powers of $\ln \omega$,
\begin{align}
&  \int_0^\infty \frac{d\omega}{\omega} \, \ln^n \left(\frac{\omega}{\mu} \right) \phi_B^+(\omega,\mu)
 \ \stackrel{n=0,1,2}{=} \
 \int_0^\infty \frac{d\omega'}{\omega'} \,  \ln^n \left(\frac{\hat\omega'}{\mu} \right) \rho_B^+(\omega',\mu)
 \cr
 = &\, e^{V} \,
  \int_0^\infty \frac{d\omega'}{\omega'}\, \ln^n \left(\frac{\hat\omega'}{\mu}\right)
  \left(\frac{\mu_0}{\hat\omega'}
 \right)^{-g}  \rho_B^+(\omega',\mu_0) \,.
\end{align}
The logarithmic moments in the dual space,
\begin{align}
L_n(\mu) \equiv
 \int_0^\infty \frac{d\omega'}{\omega'} \,  \ln^n \left(\frac{\hat\omega'}{\mu} \right) \rho_B^+(\omega',\mu)\,,
\end{align}
obey a RG equation,
\begin{align}
 \frac{d L_n(\mu)}{d\ln\mu}
 &= \Gamma_{\rm cusp}(\alpha_s) \, L_{n+1}(\mu)
 - \gamma_+(\alpha_s) L_n(\mu)
 -n \, L_{n-1}(\mu)\,,
\end{align}
which is simpler than the corresponding expressions for the logarithmic moments of $\phi_B^+(\omega)$,
see \cite{Bell:2008er}.
Its solution can be explicitly written as
\begin{align}
 L_n(\mu) &=  e^{V} 
 \sum_{m=0}^\infty \, \frac{g^m}{m!} \;
 \sum_{j=0}^n \, \frac{n!}{(n-j)!\, j!} \;
 \ln^{n-j} \left(\frac{\mu_0}{\mu}\right) L_{m+j}(\mu_0) \,.
\label{Logrel}
\end{align}
However, to make use of this relation in practice, one would have to truncate the
infinite sum, i.e.\ expand the result for $g\ll 1$. 
Evidently, the solution for the spectral function (\ref{BmesonRGE1}) -- which contains the same
information as (\ref{Logrel}) -- is much simpler and very economic.

Moreover, in terms of the spectral function $\rho_B^+(\omega',\mu)$, the 
RG-improved factorization
formulas in exclusive $B$-decays take a particularly simple form.
For example, using the results from \cite{Lunghi:2002ju,Bosch:2003fc,Beneke:2011nf},
the form factor relevant for the radiative
leptonic $B\to\gamma\ell\nu$ decay in the heavy-quark limit, can be written as
\begin{align}
 F(E_\gamma) &= H(E_\gamma,\mu) \, \int\limits_0^\infty
  \frac{d\omega'}{\omega'} \, j(2 E_\gamma \hat \omega',\, \mu) \, \rho_B^+(\omega',\mu) \,.
\end{align}
Here $H(E_\gamma,\mu)$ contains the hard-matching coefficients from QCD onto SCET and HQET, 
and satisfies the RG equation
\begin{align}
 \frac{dH(E_\gamma,\mu)}{d\ln\mu} &= \left[ \Gamma_{\rm cusp}(\alpha_s) \, \ln \frac{2E_\gamma}{\mu} - \gamma_h(\alpha_s) \right] 
 H(E_\gamma,\mu) \,,
\end{align}
and $j(s',\mu)$ is the hard-collinear function which has a
perturbative expansion
\begin{align}
 j(s',\mu) &= 1 + \frac{\alpha_s C_F}{4\pi} \left(
  \ln^2 \frac{s'}{\mu^2} -1 - \frac{\pi^2}{6} \right) + \ldots 
\end{align}
and obeys the simple RG equation
\begin{align}
 \frac{dj(s',\mu)}{d\ln\mu} &= - \bigg[ \Gamma_{\rm cusp}(\alpha_s) \,  \ln \frac{s'}{\mu^2}
 + \gamma_{\rm hc}(\alpha_s) \bigg]  \,j(s',\mu) \,,
\end{align}
with $\gamma_{\rm hc}={\cal O}(\alpha_s^2)$.
Notice that the hard-collinear function in dual space depends on $\hat \omega'$ via the combination
$
  \ln( 2 E_\gamma \hat \omega'/\mu^2)
$, where $\hat \omega'$ is defined in (\ref{hatVdef}).
The resulting form factor is RG-invariant,
$ dF(E_\gamma)/d\ln\mu = 0 \,,$
when $(\gamma_h+\gamma_{\rm hc}+\gamma_+)=0$, which has been checked at one-loop accuracy.
The RG-improved factorization formula for the form factor can
thus be written as
\begin{align}
 F(E_\gamma) &= \left[e^{V_h(\mu,\mu_h)}\left( \frac{\mu_h}{2E_\gamma} \right)^{-g(\mu,\mu_h)} 
 H(E_\gamma,\mu_h) \right] 
 \cr & \quad \times 
 \int\limits_0^\infty
  \frac{d\omega'}{\omega'} \left[ e^{-2V_{\rm hc}(\mu,\mu_{\rm hc})}
  \left( \frac{\mu_{\rm hc}^2}{2E_\gamma\hat\omega'} \right)^{g(\mu,\mu_{\rm hc})} 
  j(2E_\gamma\hat\omega',\, \mu_{\rm hc})
  \right] 
  \cr & \qquad \qquad \times
  \left[ e^{V(\mu,\mu_0)} 
  \left( \frac{\mu_0}{\hat\omega'} \right)^{-g(\mu,\mu_0)} \rho_B^+(\omega',\mu_0) \right] \,,
\end{align}
where the evolution factor for the hard coefficient, $V_h(\mu,\mu_h)$,
is defined as $V(\mu,\mu_0)$ in (\ref{RGfunctions}) with $\gamma_+$ replaced by
$\gamma_h$, and similarly for $V_{\rm hc}(\mu,\mu_{\rm hc})$ with
$\gamma_+$ replaced by $\gamma_{\rm hc}$.
Here, $\mu_{\rm hc}$ has to be chosen as a 
hard-collinear scale of order $\sqrt{2E_\gamma \langle \hat\omega'\rangle}$,
and $\mu_h$ is the hard scale of order $m_b \sim 2 E_\gamma$.\footnote{For
simplicity, we do not disentangle the matching 
and running for the $B$-meson decay constant in HQET.}


\subsection{Construction from Momentum Space}

As the second main subject of this paper,
we are now going to present a general framework to
construct momentum-space projectors from
``on-shell wave functions''
with definite relations to the
LCDAs as defined above.

\subsubsection{2-Particle Wave Functions}

An alternative method to construct momentum-space projectors that obey the
WW relations starts from a generic Dirac matrix (with the correct behaviour
under space-time transformations as
defined by the hadronic bound state)
that can be constructed from \emph{on-shell} momenta
($k^\mu$ with $k^2=0$ for the light anti-quark, and $m_b v^\mu$ with $v^2=1$ for the heavy quark),
\begin{align}
 {\cal M}_{B}^{(2)}(v,k) & = - \frac{i \tilde f_B m_B}{4} \left[ (1+\slashed v) \, \slashed k \, \gamma_5  \right]
  \, \psi_B(2 \, v \cdot k) \,,
\end{align}
and that also fulfills the free Dirac equations for both constituents,
\begin{align}
 (\slashed v - 1) \, {\cal M}_B^{(2)} = {\cal M}_B^{(2)} \, \slashed k = 0 \,.
\end{align}
The interpretation as a wave function for a 2-particle Fock state requires
to consider a particular gauge, in this case  light-cone gauge with
$n_- A(x)=0$.
We define a Lorentz-invariant integration measure $\widetilde{dk}$
for an on-shell massless particle, for which we can choose
a representation that reflects the light-cone kinematics of a hard-scattering process
(with the azimuthal angle in the transverse plane integrated out),
\begin{align}
 \widetilde{dk} & := d|k_\perp|^2 \, \frac{d\omega}{\omega} = \frac{d^3k}{\pi\, v \cdot k} \,, \qquad \mbox{with} \qquad k^\mu = \omega \,  \frac{n_+^\mu}{2} + k_\perp^\mu + \frac{|k_\perp|^2}{\omega} \, \frac{n_-^\mu}{2} \,.
\end{align}
To make contact with the general definition of LCDAs, we
consider the convolution with a hard-scattering kernel that is at most linear in $k_\perp$, and obtain
\begin{align}
 & \int \widetilde{dk} \, {\rm tr} \left[\left( T_0(\omega) + k_\perp^\mu T_\mu(\omega) \right) {\cal M}^{(2)}_B(v,k) \right]
 \cr
 = &- \frac{i \tilde f_B m_B}{4} \, {\rm tr}\left[
 \int d\omega \, T_0(\omega) \, \int d|k_\perp^2| \, \psi_B(x) \, \frac{1+\slashed v}{2} \left( \slashed n_+ + \frac{|k_\perp|^2}{\omega^2} \, \slashed n_- \right)
 \gamma_5  \right]
 \cr
 & - \frac{i \tilde f_B m_B}{4} \, {\rm tr}\left[
 \int d\omega \, T_\mu(\omega) \, \int d|k_\perp^2| \, \psi_B(x) \, \frac{1+\slashed v}{2} \left( - \frac{|k_\perp|^2}{\omega} \, \gamma_\perp^\mu \right)
 \gamma_5  \right] \,,
\end{align}
with $x= 2 \, v\cdot k = \omega+ |k_\perp|^2/\omega$.
Comparison with (\ref{Bproj}) implies
\begin{align}
 \phi_B^+(\omega) &= \int_0^\infty d|k_\perp^2| \, \psi_B(x) = \omega \, \int_\omega^\infty dx \, \psi_B(x) \,,
 \cr
 \phi_B^-(\omega) &= \int_0^\infty d|k_\perp^2| \, \frac{|k_\perp|^2}{\omega^2} \, \psi_B(x)
 = \int_\omega^\infty dx \, (x-\omega) \, \psi_B(x) \,,
 \end{align}
 together with
 \begin{align}
 \int_0^\omega d\eta \left(\phi_B^-(\eta) - \phi_B^+(\eta)\right)
 &= \omega \, \int_\omega^\infty dx \, (x-\omega) \, \psi_B(x)
 = \omega \, \phi_B^-(\omega) \,,
\end{align}
which is in line with the WW relations. Interestingly, in this approximation,
the LCDAs $\phi_B^+(\omega)$ and $\phi_B^-(\omega)$ can be obtained by simple $k_\perp$-integrals
of a light-cone wave function $\psi_B(x)$ (in particular,
this procedure could be used to match calculations in the $k_T$-factorization
approach \cite{Keum:2000wi,Lu:2000em,He:2006ud} to collinear  QCD factorization
at tree level).
Notice that in the WW approximation the wave function $\psi_B(x)$ can be reconstructed from $\phi_B^+(\omega)$ or $\phi_B^-(\omega)$ as follows,
\begin{align}
 \psi_B(x) &= \frac{\phi_B^+(x)}{x^2} - \frac{1}{x} \, \frac{d\, \phi_B^+(x)}{dx}
 = \frac{d^2\phi_B^-(x)}{dx^2} \qquad\quad \mbox{(WW)} \,.
\end{align}
This also implies that the RG evolution for $\psi_B(x,\mu)$ can be easily obtained from
that of $\rho_B^+(x,\mu)$ as in (\ref{phiBrhoB}),
\begin{align}
 \psi_B(x,\mu) &=
 \frac{1}{x} \, \int_0^\infty \frac{d\omega'}{\omega'} \,
 \frac{1}{\omega'} \, J_2\left( 2 \, \sqrt{\frac{x}{\omega'}}\right)  \rho_B^+(\omega',\mu)
 \,.
\end{align}
Using that
\begin{align}
 \rho_B^+(\omega',\mu) &=
 \int_0^\infty d\omega \, \sqrt{\frac{\omega}{\omega'}} \,
 J_1\left( 2 \, \sqrt{\frac{\omega}{\omega'} }\right)
  \, \int_{\omega}^\infty dx \, \psi_B(x,\mu)
 \cr
 &=
 \int_0^\infty dx \, x \,
 J_2\left( 2 \, \sqrt{\frac{x}{\omega'} }\right)
 \, \psi_B(x,\mu)
 \,,
\end{align}
we obtain an explicit solution for the RG evolution of the wave function
$\psi_B(x,\mu)$,
\begin{align}
\psi_B(x,\mu) &= e^V \,
 \int_0^\infty \frac{d\omega'}{\omega'} \,
  \int_0^\infty dx' \,
 \frac{x'}{x \omega'} \left( \frac{\mu_0}{\hat\omega'} \right)^{-g}
 \cr
 & \quad \times
 \, J_2\left( 2 \, \sqrt{\frac{x}{\omega'}}\right)
  J_2\left( 2 \, \sqrt{\frac{x'}{\omega'} }\right)
 \, \psi_B(x',\mu_0)
 \,,
\end{align}
which is similar to the one for the LCDA $\phi_B^+(\omega,\mu)$ in (\ref{phiBRGsol}).
We stress that the scale dependence of $\psi_B(x,\mu)$, according to our definition,
follows directly from the renormalization of the LCDAs. In the $k_T$-factorization
approach, the RG properties of the transverse-momentum-dependent wave functions
will in general be different, depending on the exact definition of the gauge-link
operator beyond the collinear limit (we refer to \cite{Collins:2003fm} for more detailed discussions).
The ansatz for the momentum-space projector is thus consistent
with the RG behaviour (within the WW approximation).

As an example, exponentially decreasing LCDAs -- which are often used in phenomenological applications
-- can be obtained from the model
\begin{align}
 \psi_B(x) & \to \ \frac{e^{-x/\omega_0}}{\omega_0^3} \quad \Leftrightarrow \quad \phi_B^+(\omega)
 \to \frac{\omega \, e^{-\omega/\omega_0}}{\omega_0^2}
 \,, \quad
 \phi_B^-(\omega) \to \frac{e^{-\omega/\omega_0}}{\omega_0}\,.
 \label{explapl}
 \end{align}
They correspond to a spectral function 
 \begin{align}
 \rho_B^+(\omega')=\rho_B^-(\omega') \to \frac{e^{-\omega_0/\omega'}}{\omega'} \,.
\end{align}
Note that the spectral function $\rho_B^+(\omega')$ shows dual behaviour compared
to the function $\phi_B^+(\omega)$, i.e.\ it is exponentially suppressed at small values of $\omega'$
and vanishes linearly with $1/\omega'$ for $\omega' \to \infty$,
whereas $\phi_B^+(\omega)$ decreases linearly at small $\omega$ and vanishes
exponentially for $\omega \to \infty$.

For comparison, a free parton picture with $v\cdot k = M_B-m_b=\bar\Lambda$ (cf.\ \cite{Kawamura:2001jm})
would correspond to
\begin{align}
  \psi_B(x) & \to \ \frac{\delta(x-2\bar\Lambda)}{2\bar\Lambda^2} \quad \Leftrightarrow \quad \phi_B^+(\omega) \to \frac{\omega}{2\bar\Lambda^2} \, \theta(2\bar\Lambda-\omega)
 \,, \quad
 \phi_B^-(\omega) \to \frac{2\bar\Lambda-\omega}{2\bar\Lambda^2} \, \theta(2\bar\Lambda-\omega)\,,
 \label{naive}
\end{align}
with a spectral function
\begin{align}
\rho_B^+(\omega')=\rho_B^-(\omega')
\to \frac{1}{{\bar \Lambda}} \, J_2\left(2 \, \sqrt {\frac{2\bar\Lambda}{\omega'}} \right)
\,.
\end{align}

\begin{figure}[t]
 \begin{center}
 \includegraphics[width=0.45\textwidth]{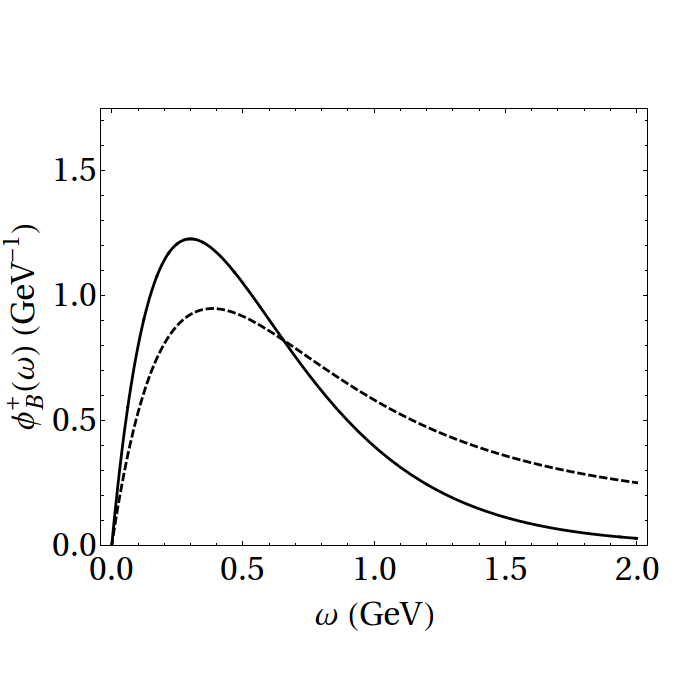} \quad \includegraphics[width=0.45\textwidth]{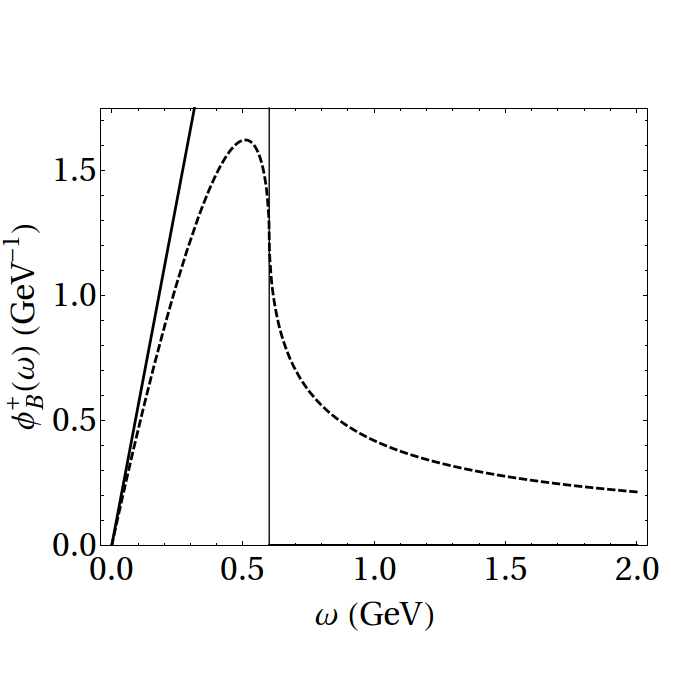}
 \\[-1.3em]
 \includegraphics[width=0.45\textwidth]{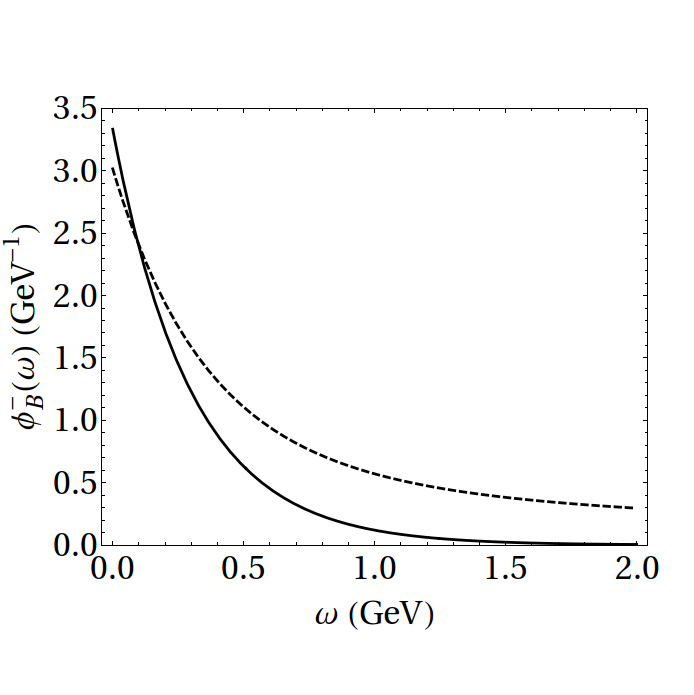} \quad \includegraphics[width=0.45\textwidth]{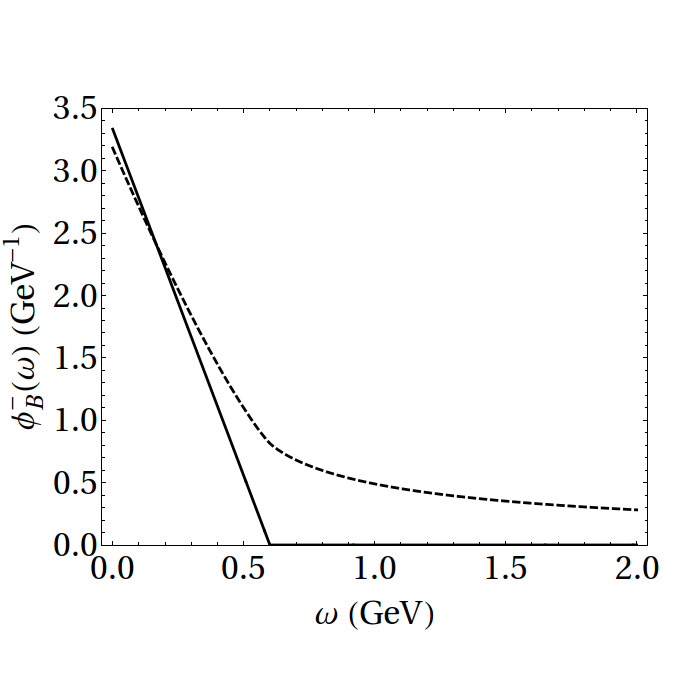}
  \\[-1.3em]
 \includegraphics[width=0.45\textwidth]{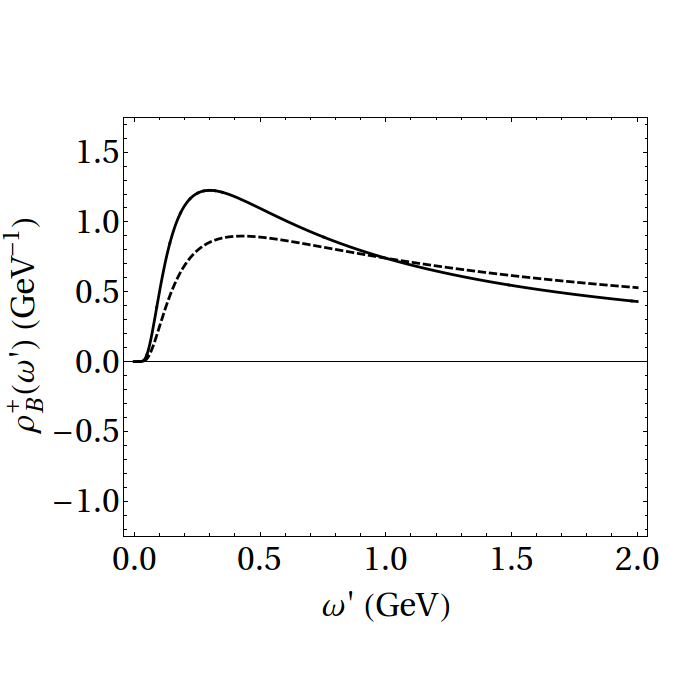} \quad \includegraphics[width=0.45\textwidth]{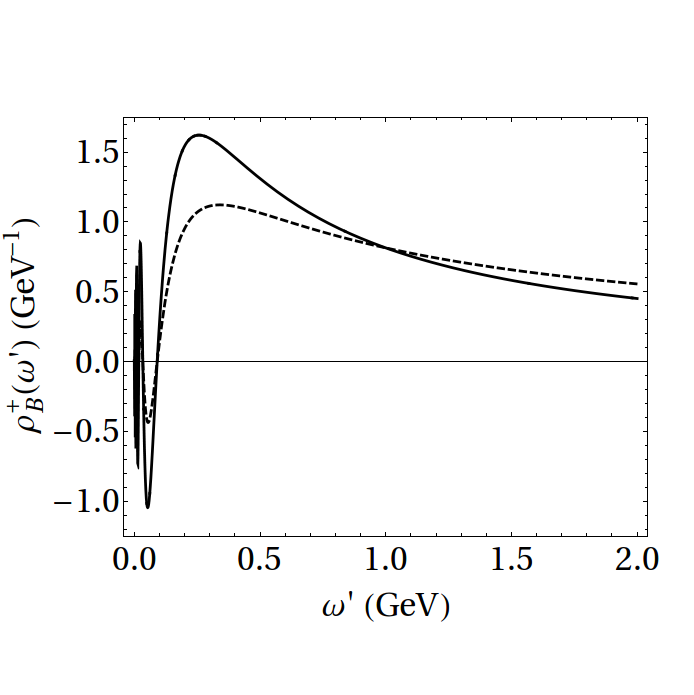}
 \end{center}
 \vspace{-0.5cm}
\caption{\label{Bmodels:plot} 
Left: exponential model from (\ref{explapl})
showing the LCDAs $\phi_B^+(\omega)$ (top), $\phi_B^-(\omega)$ (center)
and the 
spectral function $\rho_B^+(\omega')$ (bottom), using
$\mu_0=1$~GeV and $\omega_0=0.3$~GeV as input (solid line) and evolving to another scale corresponding
to $g=0.3$ (dashed line) and neglecting the 
overall 
factor $e^{V-2\gamma_E g}$ for simplicity.
Right: the same for the free parton model (\ref{naive}) with $\bar\Lambda=0.3$~GeV.
}
\end{figure}

Numerical examples for the two models with a sample RG evolution are plotted in
Fig.~\ref{Bmodels:plot}. As expected, the RG evolution tends to ``wash out'' the
differences between the shapes of the input functions at higher scales. In particular,
the LCDA of the free parton model has become a smooth function
after RG evolution (the oscillatory behaviour
of $\rho_B^+$ at small values of $\omega'$ is a relic
from the singular behaviour of $\rho_B^+$ at $\omega'=2\bar\Lambda$).

\subsubsection{Corrections to Wandzura-Wilczek Relation}

Corrections to the WW approximation can be incorporated by abandoning the
constraint ${\cal M}_B^{(2)} \slashed k =0$ from the light quark's Dirac equation.
We thus write a more general ansatz,
\begin{align}
 {\cal M}_B^{(2)}(v,k) & = - \frac{i \tilde f_B m_B}{4} \left[ (1+\slashed v)
 \left\{\slashed k \, \psi_{B1}(x) + x \, \psi_{B2}(x)  \right\}  \gamma_5  \right]
   \,.
\end{align}
Considering again the convolution with an appropriate hard-scattering kernel, one has
\begin{align}
 & \int \widetilde{dk} \, {\rm tr} \left[\left( T_0(\omega) + k_\perp^\mu T_\mu(\omega) \right) {\cal M}_B^{(2)}(v,k) \right]
 \cr
 = &- \frac{i \tilde f_B m_B}{4} \, {\rm tr}\left[
 \int d\omega \, T_0(\omega) \, \int dx \, \psi_{B1}(x) \, \frac{1+\slashed v}{2}
 \left( \omega \slashed n_+ + (x-\omega) \, \slashed n_- \right)
 \gamma_5  \right]
 \cr
  &- \frac{i \tilde f_B m_B}{4} \, {\rm tr}\left[
 \int d\omega \, T_0(\omega) \, \int dx \,  \psi_{B2}(x) \, \frac{1+\slashed v}{2}
 \left( x \, \slashed n_+ + x \, \slashed n_- \right)
 \gamma_5  \right]
 \cr
 & - \frac{i \tilde f_B m_B}{4} \, {\rm tr}\left[
 \int d\omega \, T_\mu(\omega) \, \int dx \, \psi_{B1}(x) \,
 \frac{1+\slashed v}{2} \left( -\omega \, (x-\omega) \, \gamma_\perp^\mu \right)
 \gamma_5  \right] \,.
\end{align}
The LCDAs for this ansatz follow as
\begin{align}
 \phi_B^+(\omega)
&= \int_\omega^\infty dx
 \left\{ \omega \, \psi_{B1}(x) + x \, \psi_{B2}(x) \right\} \,,
 \cr
 \phi_B^-(\omega) &=
 \int_\omega^\infty dx \,
 \left\{ (x-\omega) \, \psi_{B1}(x) + x \,  \psi_{B2}(x)  \right\}  \,,
 \end{align}
 together with
 \begin{align}
\omega \, \phi_B^-(\omega)
-  \int_0^\omega d\eta
 \left(\phi_B^-(\eta) - \phi_B^+(\eta)\right) &=
 \omega \, \int_\omega^\infty dx \,  x \,  \psi_{B2}(x)  \,.
\end{align}
The wave function $\psi_{B2}(x)$ can thus be reconstructed from
integrals of the 3-particle LCDAs $(\Psi_A-\Psi_V)$ using
(\ref{WWcorr}), and vice versa.

\subsubsection{Higher Fock States}

Our formalism can also be applied to construct on-shell wave functions for higher Fock states.
As an example, we will consider the quark-antiquark-gluon contribution in the WW approximation.
We recall that the partonic Fock-state interpretation refers to light-cone gauge
$n_- A(x) = 0$, which we will assume in the following. 
For the 3-particle contributions, it is for practical purposes often sufficient to consider
convolutions with a hard-scattering kernel that does not depend on any transverse momenta.
To keep the discussion simple, we will therefore focus
on the strict collinear limit, i.e.\ we will assume a corresponding hard-scattering kernel
without the linear terms in partonic transverse momenta.

The conventional definition of 3-particle LCDAs starts from the position-space matrix element
\cite{Kawamura:2001jm} (see also \cite{Geyer:2005fb,Huang:2005kk})
\begin{align}
 &  z_\nu \, \langle 0 | \bar q^\beta(z) \, [z,uz] \, g G^{\mu\nu}(u z) \, [uz,0]\, h_v^\alpha(0)| \bar B(v)\rangle
 \cr
 &=
  \frac{\tilde f_B m_B}{2} \left[ \frac{1+\slashed v}{2} \bigg\{
  (v^\mu \slashed z - t \gamma^\mu) \, \big( \tilde \Psi_A(t,u) - \tilde\Psi_V(t,u) \big) \right.\cr
 & \hspace{3.1cm}
 \left.-i \sigma^{\mu\nu} z_\nu \tilde \Psi_V(t,u) - z^\mu \tilde X_A(t,u)
  + \frac{z^\mu}{t} \slashed z \tilde Y_A(t,u)  \bigg\} \gamma_5 \right]^{\alpha\beta} ,
  \label{3-particleLCDA}
\end{align}
with  $t = v\cdot z$. In order to relate this representation to the wave-function approach, we need 
the momentum-space projector corresponding to the non-local matrix element
\begin{align}
\langle 0 | \bar q^\beta(z) \, g   \,A^{\mu}(uz) \, h_v^\alpha(0)| \bar B(v)\rangle
\label{3-particle matrix element}
\end{align}
in light-cone gauge. In the collinear approximation, we further have
$z^\mu = \tau n_-^\mu$, such that $t=\tau$ in a frame\footnote{In Appendix~\ref{3general}, we show
the result for a general Lorentz frame.} where $(v \cdot n_-)=1$. We then obtain
\begin{align}
& {\cal M}_B^{(3)}{}^\mu(v,\omega,\xi)
= - {i \over \xi} \, \frac{\tilde f_B m_B}{2} \left[ \frac{1+\slashed v}{2} \bigg\{
  (n_-^\mu \slashed n_- - n_-^\mu - \gamma_\perp^\mu) \big( \Psi_A(\omega,\xi) - \Psi_V(\omega,\xi) \big)
 \right.\cr
 & \hspace{1.1cm}
 \left.  +(n_-^\mu \slashed n_- - n_-^\mu + \gamma_\perp^\mu \slashed n_-)   \Psi_V(\omega,\xi)
  - n_-^\mu  X_A(\omega,\xi) + n_-^\mu \slashed n_-  Y_A(\omega,\xi)  \bigg\} \gamma_5 \right]   \,,
\label{3particleprojector}
\end{align}
where we introduced the Fourier-transformed LCDAs
\begin{align}
 \Psi_A(\omega,\xi) &= \int \frac{d\tau}{2\pi} \, \int \frac{du}{2\pi} \;  \tau\, e^{i (\omega+\xi u) \tau} \,  \tilde  \Psi_A(\tau,u)  \,,
 \qquad\text{etc.}
\end{align}
Proceeding in analogy to the 2-particle construction, we may formulate an equivalent representation of the momentum-space projector
starting from on-shell momenta for the Fock-state components
($k^\mu$ with $k^2=0$ for the light anti-quark, $l^\mu$ with $l^2=0$ for the gluon,
and $m_b v^\mu$ with $v^2=1$ for the heavy quark). The most general ansatz that fulfills 
the equations of motion for all constituents,
\begin{align}
 (\slashed v - 1) \, {\cal M}_B^{(3)}{}^\mu = {\cal M}_B^{(3)}{}^\mu \, \slashed k
 = {\cal M}_B^{(3)}{}^\mu \, l_\mu = 0\,,
\end{align}
is given by
\begin{align}
{\cal M}_B^{(3)}{}^\mu(v,k,l)
& = -i  \,\frac{\tilde f_B m_B}{2} \; (1+\slashed v) \,
\bigg\{ y k^\mu \Phi_1(x,y)
+ x l^\mu \Phi_2(x,y)
+ x\big(y \gamma^\mu - 2v^\mu \slashed l \big) \Phi_3(x,y)
\cr
&  \hspace{1.0cm}
+ k^\mu \slashed l \Phi_4(x,y)
+ l^\mu \slashed l \Phi_5(x,y)
+ x\big(\gamma^\mu \slashed l- \slashed l \gamma^\mu \big) \Phi_6(x,y)
\bigg\} \, \slashed k \gamma_5\,.
\label{3particleansatz}
\end{align}
In general, the wave functions $\Phi_i$ depend on three invariants,
$x= 2 \, v\cdot k$, $y= 2 \, v\cdot l$ and $z= 2 \, k\cdot l$, but in the above decomposition we have
neglected the invariant mass of the antiquark-gluon subsystem, $z = (k+l)^2 \simeq 0$, for simplicity,
which is based on the assumption that the wave functions only depend on the total invariant mass
of the partonic configuration, $(m_b v + k + l)^2 \simeq m_b^2 + m_b (x+y)$.
 Writing
\begin{align}
 k^\mu = \omega \,  \frac{n_+^\mu}{2} + k_\perp^\mu + \frac{|k_\perp|^2}{\omega} \, \frac{n_-^\mu}{2} \,,
 \qquad\quad
  l^\mu = \xi \,  \frac{n_+^\mu}{2} + l_\perp^\mu + \frac{|l_\perp|^2}{\xi} \, \frac{n_-^\mu}{2} \,,
\end{align}
which implies $x= \omega+ |k_\perp|^2/\omega$ and $y = \xi+ |l_\perp|^2/\xi$, we may neglect 
any odd powers of $k_\perp$ and $\l_\perp$ in the transverse-momentum integrals with a hard-scattering kernel.
In light-cone coordinates,
the momentum-space projector (\ref{3particleansatz}) may then be rewritten in terms of six independent
structures, which we choose as
\begin{align}
n_-^\mu\,,\
n_+^\mu\,,\
\gamma_\perp^\mu\,,\
n_-^\mu \slashed n_-\,,\
n_+^\mu \slashed n_-\,,\
\gamma_\perp^\mu \slashed n_-\,.
\end{align}
The light-cone gauge condition, $ {\cal M}_B^{(3)}{}^\mu \, {n_-}_\mu = 0$, further eliminates the $n_+^\mu$ and
$n_+^\mu \slashed n_-$ structures (we use these constraints to determine $\Phi_1$ and $\Phi_4$).
The remaining four Lorentz structures are of the form  (\ref{3particleprojector}), and we can
read off
\begin{align}
&\Psi_A(\omega,\xi) - \Psi_V(\omega,\xi) = \xi\,
\int_\omega^\infty dx \int_\xi^\infty dy
\bigg\{ \xi (\omega-x)x \,\Phi_2(x,y)
+ 2 \big(\omega y - \xi (\omega - x)\big) x \,\Phi_3(x,y)\cr
&\hspace{1cm}
+ \xi \big((\omega - x) \xi + (\xi - y) \omega\big) \,\Phi_5(x,y)
+ 2 \big((\omega - x) \xi + 2 (\xi - y) \omega\big) x \,\Phi_6(x,y)
\bigg\} \,, \cr
& \Psi_V(\omega,\xi) = {\xi \over 2} \,
\int_\omega^\infty dx \int_\xi^\infty dy
\bigg\{ 2 x  ( x (y-2 \xi) + 2 \omega (\xi-y) ) \,\Phi_3(x,y)  \cr
& \hspace{1cm} + \xi ( \omega ( 3 y- 4\xi) + x(3 \xi - 2 y) )\,\Phi_5(x,y)
+ 4x \omega (y-\xi )\,\Phi_6(x,y)
\bigg\} \,, \cr
&\Psi_A(\omega,\xi)  +  X_A(\omega,\xi) = \xi\,
\int_\omega^\infty dx \int_\xi^\infty dy
\bigg\{ x (\xi x -y \omega) \,\Phi_2(x,y)
- (y-\xi) (y \omega - \xi x) \,\Phi_5(x,y)\cr
&\hspace{1cm}
+ 2 x \big(\xi (x-\omega) + \omega (y - \xi)\big) \,
\left ( \Phi_3(x,y) - \Phi_6(x,y) \right )
\bigg\} \,, \cr
&\Psi_A(\omega,\xi) + Y_A(\omega,\xi) = - {\xi \over 2 \omega} \,
\int_\omega^\infty dx \int_\xi^\infty dy
\bigg\{ x (x - 2 \omega) (\xi x -y \omega) \,\Phi_2(x,y)
\cr
&\hspace{1cm}
+  \big( \xi x - y \omega \big)^2 \,\Phi_5(x,y)
- 2 x^2 \big( \xi (x-\omega) + \omega(y-\xi)   \big)  \,
\left (\Phi_3(x,y) - \Phi_6(x,y) \right )
\bigg\} \,.
\end{align}
In the collinear approximation, the four 3-particle LCDAs
$\Psi_A, \Psi_V, X_A$ and $Y_A$ can thus be expressed through four independent on-shell wave functions.

\section{$\Lambda_b$--Baryons}

\subsection{Light-Cone Distribution Amplitudes}

Light-cone distribution amplitudes for $\Lambda_b$-baryons in HQET have
been classified in \cite{Ball:2008fw} (for related work,
see also \cite{Ali:2012pn}).
They contain the hadronic information entering factorization theorems
for exclusive $\Lambda_b$ transitions in the heavy-quark limit
(see e.g.\ \cite{Feldmann:2011xf,Wang:2009hra}).
In this work, we will focus on the LCDAs related to the leading
3-particle operators \cite{Ball:2008fw} (gauge-links are understood implicitly, but not shown for simplicity),
\begin{align}
 \epsilon^{abc} \, \langle 0| \left(u^a(\tau_1 n_-) \, C\gamma_5 \, \slashed n_- \, d^b(\tau_2n_-)\right) h_v^c(0)|\Lambda_b(v,s)\rangle
 &=
 f_{\Lambda_b}^{(2)} \, \tilde\phi_2(\tau_1,\tau_2) \, u_{\Lambda_b}(v,s) \,,
 \cr
 \epsilon^{abc} \, \langle 0| \left(u^a(\tau_1 n_-) \, C\gamma_5 \, \slashed n_+ \, d^b(\tau_2n_-)\right) h_v^c(0)|\Lambda_b(v,s)\rangle
 &=
 f_{\Lambda_b}^{(2)} \, \tilde\phi_4(\tau_1,\tau_2) \, u_{\Lambda_b}(v,s) \,,
 \label{baryondef}
 \end{align}
 for the chiral-odd part (i.e.\ with an odd number of Dirac matrices in the light-diquark current),
 and
 \begin{align}
 \epsilon^{abc} \, \langle 0| \left(u^a(\tau_1 n_-) \, C\gamma_5 \,  d^b(\tau_2n_-)\right) h_v^c(0)|\Lambda_b(v,s)\rangle
 &=
 f_{\Lambda_b}^{(1)} \, \tilde\phi_3^s(\tau_1,\tau_2) \, u_{\Lambda_b}(v,s) \,,
 \cr
 \epsilon^{abc} \, \langle 0| \left(u^a(\tau_1 n_-) \, C\gamma_5 \, i\sigma_{\mu\nu} n_+^\mu n_-^\nu \, d^b(\tau_2n_-)\right) h_v^c(0)|\Lambda_b(v,s)\rangle
 &=
 2 \,f_{\Lambda_b}^{(1)} \, \tilde\phi_3^\sigma(\tau_1,\tau_2) \, u_{\Lambda_b}(v,s) \,,
\label{baryondefeven}
\end{align}
for the chiral-even part. Here $u_{\Lambda_b}(v,s)$ denotes the
on-shell Dirac spinor for the $\Lambda_b$-baryon,
and the prefactors $f_{\Lambda_b}^{(1,2)}$ are defined by the
normalization of the matrix elements of the corresponding \emph{local} operators 
in HQET at a given renormalization scale.

\subsubsection{Light-Cone Projectors for 3-Particle Fock State}

The above definitions can be cast into a manifestly Lorentz-invariant
form by defining the most general non-local matrix elements in coordinate space as \cite{Feldmann:2011xf}
\begin{align}
 & \epsilon^{abc} \, \langle 0| \left(u^a_\alpha (z_1) \,  d^b_\beta(z_2)\right) h^c_{v}(0)
  |\Lambda_b(v,s)\rangle
  \cr
 \equiv & \frac14 \left \{ f^{(1)}_{\Lambda_b} \left[ \tilde M^{(1)}(v,z_1,z_2) \gamma_5 C^T\right]_{\beta\alpha}
 +
f^{(2)}_{\Lambda_b} \left[ \tilde M^{(2)}(v,z_1,z_2) \gamma_5 C^T\right]_{\beta\alpha} \right\} u_{\Lambda_b}(v,s) \,,
\end{align}
with a part that contains an odd number of Dirac matrices ($t_i = v\cdot z_i$),
\begin{align}
\tilde M^{(2)}(v,z_1,z_2) &=
\slash v \, \tilde \Phi_2(t_1,t_2,z_1^2,z_2^2,z_1 \cdot z_2)
+
\frac{\tilde \Phi_X(t_1,t_2,z_1^2,z_2^2,z_1\cdot z_2)}{4 t_1 t_2} \left( \slash z_2 \slash v \slash z_1 - \slash z_1 \slash v \slash z_2 \right)
\cr & \quad
+
\frac{\tilde \Phi_{42}^{(i)}(t_1,t_2,z_1^2,z_2^2, z_1\cdot z_2)}{2t_1} \, \slash z_1
+
\frac{\tilde \Phi_{42}^{(ii)}(t_1,t_2,z_1^2,z_2^2,z_1 \cdot z_2)}{2t_2} \, \slash z_2 \,,
\cr
& \quad
\label{barygeneral1}
\end{align}
and a part that contains an even number of Dirac matrices,
\begin{align}
\tilde M^{(1)}(v,z_1,z_2) &=
\tilde \Phi_3^{(0)}(t_1,t_2,z_1^2,z_2^2,z_1 \cdot z_2)
+
\frac{\tilde \Phi_Y(t_1,t_2,z_1^2,z_2^2,z_1\cdot z_2)}{4 t_1 t_2}
\left( \slash z_2 \slash z_1 - \slash z_1 \slash z_2 \right)
\cr & \quad
+
\frac{\tilde \Phi_{3}^{(i)}(t_1,t_2,z_1^2,z_2^2, z_1\cdot z_2)}{2t_1} \, \slash v \slash z_1
+
\frac{\tilde \Phi_{3}^{(ii)}(t_1,t_2,z_1^2,z_2^2,z_1 \cdot z_2)}{2t_2} \, \slash z_2 \slash v \,.
\cr
& \quad
\label{barygeneral2}
\end{align}
Considering isospin invariance for the light-quark fields (exchanging $z_1 \leftrightarrow z_2$ and taking
care of the charge-conjugation properties of the Dirac matrices),
one obtains the following relations between the individual functions,
\begin{align}
\tilde\Phi_2(t_1,t_2,z_1^2,z_2^2, z_1 \cdot z_2)  &= \tilde\Phi_2(t_2,t_1,z_2^2,z_1^2, z_1 \cdot z_2) \,,
\cr
\tilde\Phi_{42}^{(i)}(t_1,t_2,z_1^2,z_2^2, z_1 \cdot z_2)& = \tilde\Phi_{42}^{(ii)}(t_2,t_1,z_2^2,z_1^2, z_1 \cdot z_2) \,,
 \cr
\tilde\Phi_X(t_1,t_2,z_1^2,z_2^2, z_1 \cdot z_2) &= \tilde\Phi_X(t_2,t_1,z_2^2,z_1^2, z_1 \cdot z_2) \,,
\end{align}
and
\begin{align}
\tilde\Phi_3^{(0)}(t_1,t_2,z_1^2,z_2^2, z_1 \cdot z_2)  &= \tilde\Phi_3^{(0)}(t_2,t_1,z_2^2,z_1^2, z_1 \cdot z_2) \,,
\cr
\tilde\Phi_{3}^{(i)}(t_1,t_2,z_1^2,z_2^2, z_1 \cdot z_2)& = \tilde\Phi_{3}^{(ii)}(t_2,t_1,z_2^2,z_1^2, z_1 \cdot z_2) \,,
 \cr
\tilde\Phi_Y(t_1,t_2,z_1^2,z_2^2, z_1 \cdot z_2) &= \tilde\Phi_Y(t_2,t_1,z_2^2,z_1^2, z_1 \cdot z_2) \,.
\end{align}

\subsubsection{The Chiral-Odd Projector $\tilde M^{(2)}$}

In the same way as we argued for $B$-mesons, we may again expand the arguments
$z_1$ and $z_2$ around the light-cone, such that $(n_-z_i)\ll z_i^\perp \ll (n_+z_i)$, to obtain
the projector in coordinate-space as
\begin{align}
\tilde M^{(2)}(v,z_1,z_2) & \longrightarrow
\frac{\slash n_+}{2} \, \tilde \phi_2(\tau_1,\tau_2)
+
\frac{\slash n_-}{2} \left( \tilde \phi_2(\tau_1,\tau_2)+\tilde \phi_{42}^{(i)}(\tau_1,\tau_2) + \tilde \phi_{42}^{(ii)}(\tau_1,\tau_2) \right)
\cr & \quad
+
\frac{\tilde \phi_{42}^{(i)}(\tau_1,\tau_2)}{2\tau_1} \,  \slash z_1^\perp
+
\frac{\tilde \phi_{42}^{(ii)}(\tau_1,\tau_2)}{2\tau_2} \,  \slash z_2^\perp
\cr
& \quad
+
\tilde\phi_X(\tau_1,\tau_2) \left( \frac{\slash z_1^\perp}{2\tau_1} - \frac{\slash z_2^\perp}{2\tau_2}
 \right) \left(\frac{\slash n_- \slash n_+}{4} - \frac{\slash n_+ \slash n_-}{4} \right)
 + {\cal O}(z_{i\perp}^2, n_- z_i) \,.
 \label{M2zexp}
\end{align}
Here again we denote with $\tau_i = \frac{n_+ z_i}{2}$
the Fourier-conjugate variables to the momentum components
$\omega_i = (n_- k_i)$ of the associated light-quark states in the heavy baryon, such that
\begin{align}
 \phi_2(\omega_1,\omega_2) & \equiv \int \frac{d\tau_1}{2\pi} \, e^{i \omega_1 \tau_1}\, \int \frac{d\tau_2}{2\pi}  \,e^{i \omega_2 \tau_2}\,
  \tilde\phi_2(\tau_1,\tau_2) \qquad \mbox{etc.}
\end{align}
Comparison with the definition in (\ref{baryondef}) yields the relation
\begin{align}
\tilde \phi_{42}^{(i)}(\tau_1,\tau_2) +\tilde \phi_{42}^{(ii)}(\tau_1,\tau_2) &=\tilde \phi_4(\tau_1,\tau_2)-\tilde\phi_2(\tau_1,\tau_2) \,,
\end{align}
while the asymmetric combination of $\tilde\phi_{42}^{(i)}$ and $\tilde\phi_{42}^{(ii)}$, as well as $\tilde\phi_X$
do not contribute in the collinear limit $z_i^2 \to 0$.
After Fourier transformation, the general momentum-space representation for (\ref{M2zexp})
including the first-order terms off the light-cone becomes
\begin{align}
M^{(2)}(\omega_1,\omega_2) = & \  \frac{\slash n_+}{2} \, \phi_2(\omega_1,\omega_2)
+
\frac{\slash n_-}{2} \, \phi_4(\omega_1,\omega_2)
\cr
&  - \frac{1}{2} \, \gamma_\mu^\perp \, \int_0^{\omega_1} d\eta_1 \left( \phi_{42}^{(i)}(\eta_1,\omega_2) - \phi_X(\eta_1,\omega_2) \right)
\frac{\slash n_+\slash n_-}{4}
\, \frac{\partial}{\partial k_{1\mu}^\perp}
\cr & - \frac{1}{2} \, \gamma_\mu^\perp \, \int_0^{\omega_1} d\eta_1 \left( \phi_{42}^{(i)}(\eta_1,\omega_2) + \phi_X(\eta_1,\omega_2) \right)
\frac{\slash n_-\slash n_+}{4}
\, \frac{\partial}{\partial k_{1\mu}^\perp}
\cr & - \frac{1}{2} \, \gamma_\mu^\perp \, \int_0^{\omega_2} d\eta_2 \left( \phi_{42}^{(ii)}(\omega_1,\eta_2) - \phi_X(\omega_1,\eta_2) \right)
\frac{\slash n_-\slash n_+}{4}
\, \frac{\partial}{\partial k_{2\mu}^\perp}
\cr & - \frac{1}{2} \, \gamma_\mu^\perp \, \int_0^{\omega_2} d\eta_2 \left( \phi_{42}^{(ii)}(\omega_1,\eta_2) + \phi_X(\omega_1,\eta_2) \right)
\frac{\slash n_+\slash n_-}{4}
\, \frac{\partial}{\partial k_{2\mu}^\perp} \,.
 \label{M2proj}
\end{align}
Compared to the mesonic analogue, we observe that a larger number of independent terms
that are sensitive to the transverse momenta of the light quarks in the hard-scattering
kernel appear.

\subsubsection{The Chiral-Even Projector $\tilde M^{(1)}$}

Similarly, for the chiral-even projector we obtain
the expansion around the collinear limit as
\begin{align}
\tilde M^{(1)}(v,z_1,z_2) & \longrightarrow
\tilde \phi_3^{(0)}(\tau_1,\tau_2)
+
\tilde\phi_{3}^{(i)}(\tau_1,\tau_2) \, \frac{\slash n_+ \slash n_-}{4}
+
\tilde\phi_{3}^{(ii)}(\tau_1,\tau_2) \, \frac{\slash n_- \slash n_+}{4}
\cr & \quad
+
\tilde \phi_{3}^{(i)}(\tau_1,\tau_2)\, \frac{\slash v \slash z_1^\perp}{2\tau_1}
+
\tilde \phi_{3}^{(ii)}(\tau_1,\tau_2) \, \frac{\slash z_2^\perp \slash v}{2\tau_2}
\cr & \quad +
\tilde \phi_Y(\tau_1,\tau_2)
\left( \frac{\slash z_{2}^{\perp} \slash n_-}{2 \tau_2} + \frac{\slash n_- \slash z_{1}^{\perp}}{2\tau_1} \right)
 + {\cal O}(z_{i\perp}^2, n_- z_i)
\,,
\end{align}
where now from the comparison with (\ref{baryondefeven}) one has the relations
\begin{align}
 \tilde \phi_3^{s}(\tau_1,\tau_2) &=
   \frac{2\tilde \phi_3^{(0)}(\tau_1,\tau_2)
   + \tilde \phi_3^{(i)}(\tau_1,\tau_2)+\tilde \phi_3^{(ii)}(\tau_1,\tau_2)}{ 2 }
   \,,
   \cr
 \tilde \phi_3^{\sigma}(\tau_1,\tau_2) &=
   \frac{\tilde \phi_3^{(ii)}(\tau_1,\tau_2)-\tilde \phi_3^{(i)}(\tau_1,\tau_2)}{ 2 }
   \,.
\end{align}
It is sometimes more convenient to define symmetric and antisymmetric
combinations of the functions $\phi_3^s$ and $\phi_3^\sigma$ which
will be denoted as \cite{Ball:2008fw}
\begin{align}
 \tilde \phi_3^{+-}(\tau_1,\tau_2) &= 2 \left(  \tilde\phi_3^s(\tau_1,\tau_2) +  \tilde\phi_3^\sigma(\tau_1,\tau_2) \right) =
 2 \left( \tilde\phi_3^{(0)}(\tau_1,\tau_2) + \tilde\phi_3^{(ii)}(\tau_1,\tau_2) \right)
 \,, \cr
  \tilde\phi_3^{-+}(\tau_1,\tau_2) &= 2 \left(  \tilde\phi_3^s(\tau_1,\tau_2) -  \tilde\phi_3^\sigma(\tau_1,\tau_2) \right) =
   2 \left(  \tilde\phi_3^{(0)}(\tau_1,\tau_2) + \tilde\phi_3^{(i)}(\tau_1,\tau_2) \right)
 \,.
\end{align}
The expansion of the corresponding momentum-space projector then takes the general form
\begin{align}
M^{(1)}(\omega_1,\omega_2) = & \  \frac{\slash n_-\slash n_+}{8} \, \phi_3^{+-}(\omega_1,\omega_2)
+ \frac{\slash n_+\slash n_-}{8} \, \phi_3^{-+}(\omega_1,\omega_2)
\cr
&  - \frac{1}{2} \, \int_0^{\omega_1} d\eta_1 \, \phi_{3}^{(i)}(\eta_1,\omega_2) \,
\slash v \, \gamma_\mu^\perp
\, \frac{\partial}{\partial k_{1\mu}^\perp}
- \frac{1}{2} \,  \int_0^{\omega_2} d\eta_2 \, \phi_{3}^{(ii)}(\omega_1,\eta_2)\,
\gamma_\mu^\perp \,\slash v
\, \frac{\partial}{\partial k_{2\mu}^\perp}
\cr & - \frac{1}{2}  \, \int_0^{\omega_1} d\eta_1 \, \phi_Y(\eta_1,\omega_2)\,
\slash n_-  \gamma_\mu^\perp
\, \frac{\partial}{\partial k_{1\mu}^\perp}
- \frac{1}{2} \, \int_0^{\omega_2} d\eta_2 \, \phi_Y(\omega_1,\eta_2)
\, \gamma_\mu^\perp  \, \slash n_-
 \frac{\partial}{\partial k_{2\mu}^\perp}
\,.
\cr &
\label{M1coordexp}
\end{align}
Again, in the general case, it involves four independent structures related
to transverse momenta of the light quarks.

\subsubsection{Wandzura-Wilczek Approximation}

In the WW approximation, the matrices $\tilde M^{(1,2)}(z_1,z_2)$ fulfill the
equations of motion for free quark fields,
\begin{align}
 \gamma_\mu \, i\partial_2^\mu \, \tilde M^{(1,2)}(v,z_1,z_2) =
 i\partial_1^\mu \, \tilde M^{(1,2)}(v,z_1,z_2) \,\gamma_\mu = 0 \,,
\end{align}
which translates into differential equations for the LCDAs in the collinear limit.
 These can be
obtained by expanding the above equations around the light-cone,
and solving for the derivatives with respect to the arguments $(z_i^2,z_1\cdot z_2)$ off the light cone.
Alternatively, one can start from the expanded form of $\tilde M^{(1,2)}$ and consider the projected equations
\begin{align}
 \frac{\slash n_+\slash n_-}{4} \, \gamma_\mu \, i\partial_2^\mu \, \tilde M^{(1,2)}(v,z_1,z_2) \big|_{z_{1,2}^\perp=0}
 = i\partial_1^\mu \, \tilde M^{(1,2)}(v,z_1,z_2) \,\gamma_\mu \frac{\slash n_-\slash n_+}{4}
 \big|_{z_{1,2}^\perp=0} = 0 \,.
\end{align}
In both cases, this yields the following WW relations for the LCDAs
in $\tilde M^{(2)}$,
\begin{align}
 \tilde\phi_{42}^{(i)}(\tau_1,\tau_2)
+ \tilde\phi_X(\tau_1,\tau_2)
+ \tau_1 \, \frac{\partial}{\partial \tau_1} \, \tilde \phi_4(\tau_1,\tau_2)
= 0 \,,
\cr
\tilde\phi_{42}^{(ii)}(\tau_1,\tau_2)
+ \tilde\phi_X(\tau_1,\tau_2)
+ \tau_2 \, \frac{\partial}{\partial \tau_2}\, \tilde \phi_4(\tau_1,\tau_2)
= 0 \,.
\end{align}
For the Fourier-transformed LCDAs this implies
\begin{align}
 \phi_{42}^{(i)}(\omega_1,\omega_2)
+ \phi_X(\omega_1,\omega_2)
- \frac{\partial}{\partial\omega_1} \left( \omega_1 \, \phi_4(\omega_1,\omega_2) \right)
= 0 \,,
\cr
 \phi_{42}^{(ii)}(\omega_1,\omega_2)
+ \phi_X(\omega_1,\omega_2)
- \frac{\partial}{\partial\omega_2} \left( \omega_2 \, \phi_4(\omega_1,\omega_2) \right)
= 0 \,.
\end{align}
Equivalently, by considering linear combinations of the above, the
following relations hold
\begin{align}
 \phi_{42}^{(i)}(\omega_1,\omega_2) - \phi_{42}^{(ii)}(\omega_1,\omega_2) &=
 \frac{\partial}{\partial\omega_1} \left( \omega_1 \, \phi_4(\omega_1,\omega_2) \right)
 - \frac{\partial}{\partial\omega_2} \left( \omega_2 \, \phi_4(\omega_1,\omega_2) \right) \,,
 \cr
 2 \,\phi_X(\omega_1,\omega_2)+ \phi_4(\omega_1,\omega_2) - \phi_2(\omega_1,\omega_2)  & =
 \frac{\partial}{\partial\omega_1} \left( \omega_1 \, \phi_4(\omega_1,\omega_2) \right)
 + \frac{\partial}{\partial\omega_2} \left( \omega_2 \, \phi_4(\omega_1,\omega_2) \right)\,.
\end{align}
This reveals that, once the functions $\phi_2$ and $\phi_4$ -- which are the relevant
LCDAs in the collinear limit -- are given,
the function $\phi_X$ and the asymmetric combination of $\phi_{42}^{(i,ii)}$ can be calculated
from the WW approximation.

In a similar way, for the LCDAs in $\tilde M^{(1)}$ we obtain the relations
\begin{align}
 \tilde \phi_3^{(i)}(\tau_1,\tau_2) + \tau_1 \, \frac{\partial}{\partial \tau_1}
  \left( \tilde \phi_3^{(0)}(\tau_1,\tau_2) + \tilde \phi_3^{(i)}(\tau_1,\tau_2) \right) &= 0 \,,
  \cr
 \tilde \phi_3^{(ii)}(\tau_1,\tau_2) + \tau_2 \, \frac{\partial}{\partial \tau_2}
  \left( \tilde \phi_3^{(0)}(\tau_1,\tau_2) + \tilde \phi_3^{(ii)}(\tau_1,\tau_2) \right) &= 0 \,,
\end{align}
or, in momentum space,
\begin{align}
 \phi_3^{(i)}(\omega_1,\omega_2) - \frac{\partial}{\partial \omega_1}
  \left( \omega_1 \, \phi_3^{(0)}(\omega_1,\omega_2) + \omega_1 \, \phi_3^{(i)}(\omega_1,\omega_2) \right) &= 0 \,,
  \cr
  \phi_3^{(ii)}(\omega_1,\omega_2) - \frac{\partial}{\partial \omega_2}
  \left( \omega_2 \, \phi_3^{(0)}(\omega_1,\omega_2) + \omega_2 \, \phi_3^{(ii)}(\omega_1,\omega_2) \right) &= 0
  \,.
\end{align}
Notice that in this case, the function $\phi_Y$ does not appear in the WW relations,
and therefore remains independent, whereas the functions $\phi_3^{+-}$ 
and $\phi_3^{-+}$
that are relevant in the collinear limit are related by
\begin{align}
 -\omega_1 \, \frac{\partial}{\partial \omega_1} \,\phi_3^{-+}(\omega_1,\omega_2) &=  -\omega_2 \, \frac{\partial}{\partial \omega_2} \,\phi_3^{+-}(\omega_1,\omega_2)
  =  2 \, \phi_3^{(0)}(\omega_1,\omega_2) \,.
\end{align}

\subsection{Construction from Momentum Space}

We are now going to apply the same formalism that we have developed for $B$-mesons
to construct momentum-space projectors for  $\Lambda_b$-baryons from 3-particle
wave functions.
To keep the discussion simple, we will ignore
corrections to the WW relation
in the rest of the paper.
The most general form of the momentum-space projectors can then be written as
\begin{align}
 M^{(1)}(v,k_1,k_2) &= \tilde \psi_s(x_1,x_2, K^2 ) \, \slash k_2 \, \slash k_1\,,
\qquad
M^{(2)}(v,k_1,k_2) = \tilde \psi_v(x_1,x_2, K^2) \, \slash k_2 \, \slash v \, \slash k_1 \,,
\end{align}
with $x_i = 2 \, v\cdot k_i$ and $K^2=(k_1+k_2)^2$
and two independent wave functions $\psi_s$ and $\psi_v$.
The equations of motion, $\slash k_2 \,  M^{(1,2)}(v,k_1,k_2) =  M^{(1,2)}(v,k_1,k_2)\slash k_1 =0$,
are again trivially fulfilled for on-shell quarks, $k_i^2=0$.
On the other hand, the invariant mass of the diquark system -- in principle --
can be arbitrary, $K^2 \neq 0$. For simplicity, we will ignore the potential $K^2$ dependence,
which would correspond to the case where the wave function only depends on the \emph{total}
invariant mass of the three quarks in the $\Lambda_b$-baryon,
$(m_b v + k_1 + k_2)^2 \simeq m_b^2 + m_b (x_1 + x_2)$.

\subsubsection{The Chiral-Odd Projector $M^{(2)}$}

To compare  with the general definition of LCDAs, we again
consider the convolution with a hard-scattering kernel that is at most linear in $k_{i\perp}$.
For the chiral-odd projector $M^{(2)}$, we obtain
\begin{align}
 & \int \widetilde{dk_1} \, \int \widetilde{dk_2} \,
{\rm tr}\left[\left( T_0(\omega_1,\omega_2) + k_{i\perp}^\mu T_\mu^i(\omega_1,\omega_2) \right)
 M^{(2)}(v,k_1,k_2) \right]
 \cr = &  \int d\omega_1 \, d\omega_2 \, \int\limits_{\omega_1}^\infty dx_1 \int\limits_{\omega_2}^\infty dx_2  \Bigg\{
 \cr
  & \quad
\, {\rm tr} \left[T_0(\omega_1,\omega_2) \left(
  \omega_1 \omega_2 \, \frac{\slash n_+}{2} + (x_1-\omega_1)(x_2-\omega_2) \, \frac{\slash n_-}{2} \right)
 \right]  \cr
 & -
 {\rm tr} \left[ T_\mu^1(\omega_1,\omega_2)
 \left( \omega_1 \omega_2  (x_1-\omega_1) \, \frac{\slash n_+\slash n_-}{4}
 + \omega_1 (x_1-\omega_1)(x_2-\omega_2) \, \frac{\slash n_-\slash n_+}{4} \right) \frac{\gamma_\perp^\mu}{2}
 \right]  \cr
 & -
 {\rm tr} \left[ T_\mu^2(\omega_1,\omega_2) \,\frac{\gamma_\perp^\mu}{2} \left( \omega_1 \omega_2  (x_2-\omega_2) \, \frac{\slash n_-\slash n_+}{4}
 + \omega_2 (x_1-\omega_1)(x_2-\omega_2) \, \frac{\slash n_+\slash n_-}{4} \right)
 \right]  \cr
& \qquad \Bigg\} \, \psi_v (x_1,x_2)
 \,,
\end{align}
where the momentum integrations for the light quarks are defined as in the mesonic case.
Comparison with the momentum-space projector shown in (\ref{M2proj}) above, yields
\begin{align}
 \phi_2(\omega_1,\omega_2) &=
 \int\limits_{\omega_1}^\infty dx_1 \int\limits_{\omega_2}^\infty dx_2 \,\omega_1 \omega_2 \,
   \psi_v(x_1,x_2) \,,
   \label{phi2frompsi}
  \cr
 \phi_4(\omega_1,\omega_2)&=
 \int\limits_{\omega_1}^\infty dx_1 \int\limits_{\omega_2}^\infty dx_2 \,
  (x_1-\omega_1)(x_2-\omega_2) \, \psi_v(x_1,x_2) \,,
\end{align}
together with
\begin{align}
 \phi_{42}^{(i)}(\omega_1,\omega_2)&= \frac12 \, \int_{\omega_1}^\infty dx_1 \int_{\omega_2}^\infty dx_2 \,
   x_2 \, (x_1-2\omega_1) \, \psi_v(x_1,x_2) \,,
   \cr
 \phi_{42}^{(ii)}(\omega_1,\omega_2)&= \frac12 \, \int_{\omega_1}^\infty dx_1 \int_{\omega_2}^\infty dx_2 \,
   x_1 \, (x_2-2\omega_2) \, \psi_v(x_1,x_2) \,,
\end{align}
and
\begin{align}
  \phi_X(\omega_1,\omega_2)&= \frac12 \, \int_{\omega_1}^\infty dx_1 \int_{\omega_2}^\infty dx_2 \,
    (x_1-2\omega_1)\, (x_2-2\omega_2) \, \psi_v(x_1,x_2) \,.
\end{align}
It can easily be checked  that the LCDAs constructed in this way satisfy
the WW relations as derived above.
Notice that our simplified ansatz relates \emph{all} LCDAs to $x_i$ moments of only two
fundamental wave functions $\psi_v$ (and $\psi_s$ below).
The functional form of $\psi_v$ can be reconstructed, for instance, from
\begin{align}
 \psi_v(x_1,x_2) &= \frac{d^2}{dx_1 \, dx_2} \left( \frac{\phi_2(x_1,x_2)}{x_1x_2} \right)
  =  \frac{d^4 \phi_4(x_1,x_2)}{dx_1^2 \, dx_2^2}
  \qquad\quad \mbox{(WW)} \,.
\end{align}
In a more general ansatz, these
relations would be modified by the non-trivial $K^2$-dependence of the wave functions.

In the simplest case, we could again model the wave functions by assuming an exponential dependence of $\psi_v$
on $(x_1+x_2)$ with a single hadronic parameter $\omega_0$ measuring the average energy of the
light quarks,
\begin{align}
 \psi_v(x_1,x_2) & \to \frac{\exp \left( - \frac{x_1+x_2}{\omega_0} \right)}{\omega_0^6} \,.
\end{align}
This ansatz yields the following exponential model for the various LCDAs defined above,
\begin{align}
 \phi_2(\omega_1,\omega_2) &\to \frac{\omega_1\omega_2}{\omega_0^4} \, e^{-(\omega_1+\omega_2)/\omega_0} \,,
\qquad
 \phi_4(\omega_1,\omega_2) \to \frac{1}{\omega_0^2} \, e^{-(\omega_1+\omega_2)/\omega_0} \,,
\label{M2expresults}
\end{align}
and
\begin{align}
 \phi_{42}^{(i)}(\omega_1,\omega_2)&\to
 \frac{(\omega_0-\omega_1)(\omega_0+\omega_2)}{2\omega_0^4} \, e^{-(\omega_1+\omega_2)/\omega_0} \,,
 \cr
 \phi_{42}^{(ii)}(\omega_1,\omega_2)&\to
 \frac{(\omega_0+\omega_1)(\omega_0-\omega_2)}{2\omega_0^4} \, e^{-(\omega_1+\omega_2)/\omega_0} \,,
\end{align}
and
\begin{align}
 \phi_X(\omega_1,\omega_2)&\to \frac{(\omega_0-\omega_1)(\omega_0-\omega_2)}{2\omega_0^4} \, e^{-(\omega_1+\omega_2)/\omega_0} \,.
\end{align}
with
\begin{align}
 \phi_{42}^{(i)}(\omega_1,\omega_2)-
 \phi_{42}^{(ii)}(\omega_1,\omega_2)&\to
 \frac{\omega_2-\omega_1}{\omega_0^3} \, e^{-(\omega_1+\omega_2)/\omega_0} \,.
\end{align}

For comparison, an alternative model based on a free parton picture with
the constraint $x_1+x_2=2\bar\Lambda =M_{\Lambda_b}- m_b$ would
correspond to a wave function
\begin{align}
 \psi_v(x_1,x_2) & \to \frac{15}{4\,\bar \Lambda^5} \, \delta(x_1+x_2-2\bar\Lambda) \,.
\end{align}
From this, our construction immediately yield the corresponding expressions
for the LCDAs in terms of $\theta$-functions,
\begin{align}
 \phi_2(\omega_1,\omega_2) &\to \frac{15 \, \omega_1 \, \omega_2 \, (2\bar \Lambda- \omega_1-\omega_2)}{4\bar\Lambda^5} \,
  \theta(2\bar\Lambda-\omega_1-\omega_2) \,,
\cr
 \phi_4(\omega_1,\omega_2) &\to \frac{5 \,  (2\bar \Lambda- \omega_1-\omega_2)^3}{8\bar\Lambda^5} \,
  \theta(2\bar\Lambda-\omega_1-\omega_2)\,, \qquad \mbox{etc.}
  \label{M2partonresults}
\end{align}

\begin{figure}[t]
 \begin{center}
 \includegraphics[width=0.48\textwidth]{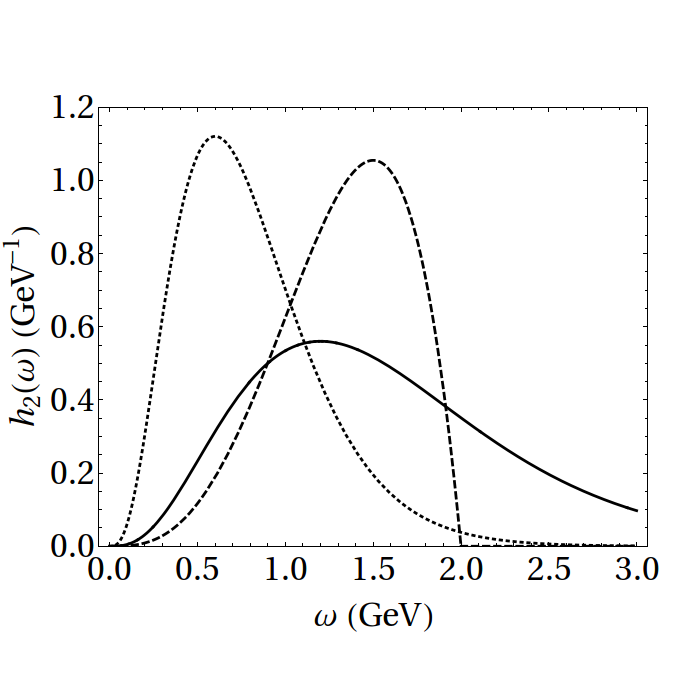}\quad
 \includegraphics[width=0.48\textwidth]{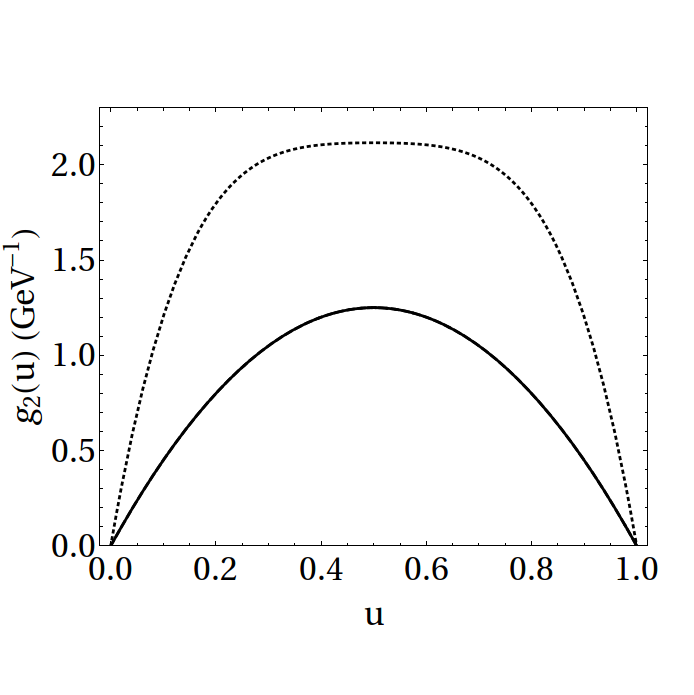}
 \end{center}
\caption{\label{f2g2} The functions $h_2(\omega)$ (left) and $g_2(u)$ (right)
for different models of the LCDA $\phi_2(\omega_1,\omega_2)$.
Solid lines correspond to the model (\ref{M2expresults}), dashed lines to
(\ref{M2partonresults}) and light-dotted lines to \cite{Ball:2008fw}
(in the right figure the dashed line is on top of the solid line).
}
\end{figure}

To illustrate these model results,
we compare the LCDA $\phi_2(\omega_1,\omega_2)$ following
from the exponential ansatz in (\ref{M2expresults}), the free-parton approximation (\ref{M2partonresults})
and the model from
Eq.~(38) in \cite{Ball:2008fw}. For that purpose,
we disentangle the dependence on the total light-cone momentum $\omega$ and
the momentum fractions $u$ of the light quarks by considering the projections
\begin{align}
 h_2(\omega) &:= \omega \,\int_0^1 du \, \phi_2(u\omega,\bar u\omega) =
 \left\{\begin{array}{ccl}
  \frac{\omega^3}{6\omega_0^4} \, e^{-\omega/\omega_0} && \mbox{(\ref{M2expresults}) with $\omega_0=\frac{2\bar\Lambda}{5}=0.4$~GeV}\\[0.2em]
\frac{\omega^3}{6\epsilon_0^4} \, e^{-\omega/\epsilon_0} && \mbox{\cite{Ball:2008fw} with $\epsilon_0=0.2$~GeV}\\[0.2em]
            \frac{5 \omega^3 \, (2\bar\Lambda-\omega)}{8\bar\Lambda^5} \, \theta(2\bar\Lambda-\omega)
                     && \mbox{(\ref{M2partonresults}) with $\bar\Lambda=1$~GeV}
                                \end{array}
                     \right.
\label{Lambdabmodels}
\end{align}
and
\begin{align}
 g_2(u) &:= \int_0^\infty d\omega \, \phi_2(u\omega,\bar u\omega) =
 \left\{
 \begin{array}{ccl}
  \frac{2 u\bar u}{\omega_0} &&
  \mbox{(\ref{M2expresults}) with $\omega_0=0.4$~GeV}
  \\[0.2em]
  u\bar u \left(\frac{2}{\epsilon_0} + \frac{3a_2\left(5 (u-\bar u)^2-1 \right)}{\epsilon_1}  \right) &&
  \mbox{\cite{Ball:2008fw} with $\left\{
  \begin{array}{l} \epsilon_0=0.2~{\rm GeV} \\ \epsilon_1=0.65~{\rm GeV} \\ a_2=1/3 \end{array} \right.$ }\\[0.2em]
            \frac{5 u\bar u}{\bar\Lambda}
                     && \mbox{(\ref{M2partonresults}) with $\bar\Lambda=1$~GeV}
                                \end{array}
                     \right.
\end{align}
The parameter $\omega_0$ in the first case has been related to the value of $\bar\Lambda$ in the third case,
such that the $\langle \omega^{-1}\rangle$ moment of $h_2$ is identical in both cases. 
Notice that the two models for the LCDA $\phi_2$
which are based on a wave function $\psi_v$ that only depends on the \emph{sum} of
the light-quark energies, lead to 
\begin{align}
 \psi_v=\psi_v(x_1+x_2) &\quad
\Leftrightarrow \quad
 \phi_2(u\omega,\bar u\omega)\propto u\bar u \,.
\label{wouldbeasymptotic}
 \end{align}
In contrast, the model in \cite{Ball:2008fw}
takes into account a  non-trivial shape
from the next-to-leading term in a Gegenbauer expansion.
That model also 
prefers a smaller value for the  parameter $\epsilon_0$
and a corresponding larger value for the inverse moment $\langle \omega^{-1}\rangle$
than in the other two models. The numerical comparison between the three models 
is shown in Fig.~\ref{f2g2}.

\subsubsection{The Chiral-Even Projector $M^{(1)}$}

For the chiral-even projector $M^{(1)}$, we again consider the convolution with a hard-scattering kernel, and obtain
\begin{align}
 & \int \widetilde{dk_1} \, \int \widetilde{dk_2} \,
{\rm tr}\left[\left( T_0(\omega_1,\omega_2) + k_{i\perp}^\mu T_\mu^i(\omega_1,\omega_2) \right)
 M^{(1)}(v,k_1,k_2) \right]
 \cr = &  \int d\omega_1 \, d\omega_2 \, \int\limits_{\omega_1}^\infty dx_1 \int\limits_{\omega_2}^\infty dx_2  \Bigg\{
 \cr
  & \quad
\, {\rm tr} \left[  T_0(\omega_1,\omega_2) \left( \omega_2 \, (x_1-\omega_1)\,\frac{\slash n_+ \slash n_-}{4}
+  \omega_1 \,(x_2-\omega_2)\, \frac{\slash n_-\slash n_+}{4}\right)
 \right]  \cr
 & -
 {\rm tr} \left[  T_\mu^1(\omega_1,\omega_2) \left( \omega_1  \omega_2 \, (x_1-\omega_1) \,  \frac{\slash n_+}{2}
                       + \omega_1 \, (x_1-\omega_1) \, (x_2-\omega_2) \,  \frac{\slash n_-}{2} \right) \frac{\gamma_\perp^\mu}{2}
 \right]  \cr
 & -
 {\rm tr} \left[  T_\mu^2(\omega_1,\omega_2) \,\frac{\gamma_\perp^\mu}{2} \left(\omega_1  \omega_2 \, (x_2-\omega_2) \,  \frac{\slash n_+}{2}
                       + \omega_2 \, (x_1-\omega_1) \, (x_2-\omega_2) \,  \frac{\slash n_-}{2}
 \right)
 \right]  \cr
& \qquad \Bigg\} \, \psi_s (x_1,x_2)
 \,.
\end{align}
Comparison with the momentum-space expression (\ref{M1coordexp}) yields
\begin{align}
 \phi_3^{-+}(\omega_1,\omega_2) = 2 \left( \phi_3^{(0)}(\omega_1,\omega_2)+\phi_3^{(i)}(\omega_1,\omega_2)\right)
 &= 2 \,
 \int\limits_{\omega_1}^\infty dx_1 \int\limits_{\omega_2}^\infty dx_2 \,
  \omega_2 \, (x_1 - \omega_1) \, \psi_s(x_1,x_2) \,,
  \cr
 \phi_3^{+-}(\omega_1,\omega_2) = 2 \left(\phi_3^{(0)}(\omega_1,\omega_2)+\phi_3^{(ii)}(\omega_1,\omega_2)\right) &=
 2 \,
 \int\limits_{\omega_1}^\infty dx_1 \int\limits_{\omega_2}^\infty dx_2 \,
  \omega_1 \, (x_2 - \omega_2) \, \psi_s(x_1,x_2) \,,
\end{align}
and
\begin{align}
\phi_3^{(0)}(\omega_1,\omega_2) &=  \int\limits_{\omega_1}^\infty dx_1 \int\limits_{\omega_2}^\infty dx_2 \, \omega_1 \omega_2
 \, \psi_s(x_1,x_2) \,,
\cr
\phi_Y(\omega_1,\omega_2) &= \frac12 \, \int\limits_{\omega_1}^\infty dx_1 \int\limits_{\omega_2}^\infty dx_2 \, (2\omega_1-x_1) \, (2\omega_2-x_2)
 \, \psi_s(x_1,x_2)\,.
\end{align}
Again, the wave function $\psi_s$ in our approximation can be reconstructed from
\begin{align}
 \psi_s(x_1,x_2) &= \frac{d^2}{dx_1 \, dx_2} \left( \frac{\phi_3^{(0)}(x_1,x_2)}{x_1x_2} \right)
   \qquad\quad \mbox{(WW)} \,.
\end{align}
With the exponential model for the wave function, we now obtain
\begin{align}
 \psi_s(x_1,x_2) & \to \frac{\exp \left( - \frac{x_1+x_2}{\omega_0} \right)}{\omega_0^6} \,,
\end{align}
which yields
\begin{align}
\phi_3^{-+}(\omega_1,\omega_2) &\to \frac{2 \, \omega_2}{\omega_0^3} \, e^{-(\omega_1+\omega_2)/\omega_0} \,,
\qquad
\phi_3^{+-}(\omega_1,\omega_2) \to \frac{2 \,\omega_1}{\omega_0^3} \, e^{-(\omega_1+\omega_2)/\omega_0} \,,
\end{align}
and
\begin{align}
\phi_3^{(0)}(\omega_1,\omega_2) &\to \frac{\omega_1\omega_2}{\omega_0^4} \, e^{-(\omega_1+\omega_2)/\omega_0} \,,
\cr
\phi_Y(\omega_1,\omega_2) &\to \frac{(\omega_1-\omega_0)(\omega_2-\omega_0)}{2\omega_0^4} \, e^{-(\omega_1+\omega_2)/\omega_0} \,.
\end{align}
The free parton picture now yields
\begin{align}
 \phi_3^{-+}(\omega_1,\omega_2) & \to
 \frac{15 \, \omega_2 \, (2\bar\Lambda-\omega_1-\omega_2)^2}{4 \,\bar \Lambda^5} \, \theta(2\bar\Lambda- \omega_1-\omega_2)
 \,,
 \cr
 \phi_3^{+-}(\omega_1,\omega_2) & \to
 \frac{15 \, \omega_1 \, (2\bar\Lambda-\omega_1-\omega_2)^2}{4\,\bar \Lambda^5} \, \theta(2\bar\Lambda- \omega_1-\omega_2)
 \,,
\end{align}
etc.

\subsection{Renormalization-Group Evolution}

In the following, we will focus on the twist-2 $\Lambda_b$
LCDA $\phi_2(\omega_1,\omega_2)$ which enters the leading terms in factorization
theorems for exclusive heavy-to-light decay amplitudes in the heavy-quark limit,
see e.g.\ \cite{Wang:2011uv}.
Its one-loop RG equation has  been extensively discussed
in \cite{Ball:2008fw}, and reads\footnote{With a slight abuse of notation, we write a colour factor
$C_F$ in the baryon case, although more precisely, the colour factor arises as $1+1/N_C$ which
only coincides with $C_F$ for $N_C=3$. It should be noted, however, that in the LN kernel for 
$N_C\neq 3$ one has to add up the contributions from $(N_C-1)$ light spectators in a color singlet
baryon, such that the net result in front of $\Gamma_{\rm cusp} \, \ln\mu$ 
would be proportional to $C_F$ again.}
\begin{align}
 \frac{d\phi_2(\omega_1,\omega_2,\mu)}{d\ln\mu} &=
 - \left[\Gamma_{\rm cusp}(\alpha_s) \,
  \ln \frac{\mu}{\sqrt{\omega_1\omega_2}} +\gamma_+(\alpha_s) \right] \phi_2(\omega_1,\omega_2,\mu)
 \cr
 & \qquad
- \frac{\omega_1}{2} \, \int_0^\infty d\eta_1 \, \Gamma_{+}(\omega_1,\eta_1,\alpha_s) \, \phi_2(\eta_1,\omega_2,\mu)
\cr & \qquad
- \frac{\omega_2}{2} \, \int_0^\infty d\eta_2 \, \Gamma_{+}(\omega_2,\eta_2,\alpha_s) \, \phi_2(\omega_1,\eta_2,\mu)
 \cr & \qquad
   + \frac{\alpha_s C_F}{2\pi} \, \int_0^1 dv \, V^{\rm ERBL}(u,v) \, \phi_2(v\omega,\bar v\omega,\mu) \,,
\label{baryon-one-loop}
\end{align}
where $\omega=\omega_1+\omega_2$, and $u=1-\bar u=\omega_1/\omega$, and $\bar v=1-v$.
Here the first three lines correspond to the LN kernel for heavy baryons
with the same anomalous dimensions as in (\ref{one-loop}), 
whereas the last term is the ERBL kernel
\cite{Efremov:1979qk,Lepage:1979zb},
which arises from gluon exchange among
the light quarks in the heavy baryon.

\subsubsection{Analytic Solution}

We follow a similar strategy as for the $B$-meson LCDA $\phi_+$, 
and as a first step introduce the logarithmic Fourier transform
(which is in almost one-to-one correspondence
to the Mellin moments discussed in \cite{Ball:2008fw}),
\begin{align}
 \varphi_2(\theta_1,\theta_2,\mu) &=  \int_0^\infty \frac{d\omega_1}{\omega_1} \,
 \int_0^\infty \frac{d\omega_2}{\omega_2} \, \phi_2(\omega_1,\omega_2,\mu)
 \left( \frac{\omega_1}{\mu} \right)^{-i\theta_1} \left( \frac{\omega_2}{\mu} \right)^{-i\theta_2} \,.
\end{align}
Next, we introduce  the ansatz
\begin{align}
 \varphi_2(\theta_1,\theta_2,\mu) &:=
 \frac{\Gamma(1-i\theta_1) \, \Gamma(1-i\theta_2)}{\Gamma(1+i\theta_1) \, \Gamma(1+i\theta_2)}
 \, \int_0^\infty \frac{d\omega_1'}{\omega_1'} \,
 \int_0^\infty \frac{d\omega_2'}{\omega_2'} \, \rho_2(\omega_1',\omega_2',\mu)
 \left(\frac{\mu}{\omega_1'} \right)^{i\theta_1}
 \left(\frac{\mu}{\omega_2'} \right)^{i\theta_2}
\end{align}
in complete analogy
to the mesonic case, such that
\begin{align}
 \phi_2(\omega_1,\omega_2,\mu) &=
 \int_0^\infty \frac{d\omega_1'}{\omega_1'} \,\int_0^\infty \frac{d\omega_2'}{\omega_2'}
 \, \sqrt{\frac{\omega_1 \omega_2}{\omega_1' \omega_2'}} \,
 J_1\left( 2 \sqrt{\frac{\omega_1}{\omega_1'}}\right)
 J_1\left( 2 \sqrt{\frac{\omega_2}{\omega_2'}}\right)
 \rho_2(\omega_1',\omega_2',\mu) \,.
 \end{align}
The inverse transformation that expresses the dual spectral function 
$\rho_2(\omega_1',\omega_2')$
in terms of the momentum-space LCDA is then given by
\begin{align}
 \rho_2(\omega_1^{\prime},\omega_2^{\prime},\mu)=\int_0^{\infty} { d 
 \omega_1 \over \omega_1}
 \int_0^{\infty} { d \omega_2 \over \omega_2}
 \sqrt{{\omega_1 \omega_2 \over \omega_1^{\prime} \omega_2^{\prime} 
 }} \,\,
 J_1\left(2 \sqrt{{\omega_1  \over \omega_1^{\prime}}}\right) \,
 J_1\left(2 \sqrt{{\omega_2  \over \omega_2^{\prime}}}\right)  \,\,
 \phi_2(\omega_1,\omega_2,\mu)\,.
 \end{align}
The one-loop RG equation (\ref{baryon-one-loop})
can be rewritten for the spectral function $\rho_2(\omega_1',\omega_2')$
in a straightforward manner.
In the absence of the ERBL kernel, the LN terms alone would
take an analogous factorized form as in the case of the $B$-meson spectral function
$\rho_B^+(\omega',\mu)$,
\begin{align}
 \frac{d\rho_2(\omega_1',\omega_2',\mu)}{d\ln\mu}\Big|_{\rm LN} = &
 - \left[\Gamma_{\rm cusp}(\alpha_s) \, \ln \frac{\mu}{\sqrt{\hat \omega_1' \hat\omega_2'}} + \gamma_+(\alpha_s) \right]
 \rho_2(\omega_1',\omega_2',\mu) \,.
\end{align}
In this approximation the RG equation would simply be solved by
\begin{align}
 \rho_2(\omega_1',\omega_2',\mu) \Big|_{\rm LN}&=
 e^{V} \left(\frac{\mu_0}{\sqrt{\hat\omega_1' \hat\omega_2'}} \right)^{-g}
  \rho_2(\omega_1',\omega_2',\mu_0) \,,
  \label{rho2model}
\end{align}
with $\hat \omega'_i = e^{-2\gamma_E} \, \omega'_i$, and the RG functions $V$ and $g$ from
(\ref{RGfunctions}).
The derivation of the ERBL term for the evolution of $\rho_2(\omega_1',\omega_2',\mu)$, however, is more complicated
(the details can be found in Appendix~\ref{app:ERBL}). Interestingly, the final result takes a simple form
when written in terms of the reduced dual momentum and dual momentum fractions,
\begin{align} \omega_r' \equiv \frac{\omega_1' \omega_2'}{\omega_1'+\omega_2'} & \qquad \mbox{and}
\quad u'=1-\bar u' = \frac{\omega_1'}{\omega_1'+\omega_2'} \,.
\end{align}
Writing
\begin{align}
  \rho_2(\omega_1',\omega_2')\equiv \hat \rho_2(\omega_r',u') \,
\end{align}
we obtain
\begin{align}
\frac{d\hat\rho_2(\omega_r',u',\mu)}{d\ln\mu}\Big|_{\rm LN+ERBL} = &
- \left[\Gamma_{\rm cusp}(\alpha_s) \, \ln \frac{\mu \,\sqrt{u'\bar u'}}{\hat\omega_r'} + \gamma_+(\alpha_s) \right]
 \hat\rho_2(\omega_r',u',\mu)
 \cr
 & \quad + \frac{\alpha_s C_F}{2\pi} \, \int_0^1 dv' \,
  V^{\rm ERBL}(u',v') \, \hat \rho_2\left(\omega_r',v',\mu\right)
\,.\label{LNplusERBL}
\end{align}
If we expand the spectral function in terms of Gegenbauer polynomials $C_n^{(3/2)}(2u'-1)$, which are
the eigenfunctions of the ERBL kernel,
\begin{align}
 \hat\rho_2(\omega_r',u',\mu) &:= \sum_{n=0,2,4,\ldots}^\infty  u' \, \bar u' \,f_n(\omega_r',\mu) \, C_{n}^{(3/2)}(2u'-1) \,,
\end{align}
the coefficients $f_n(\omega_r',\mu)$ satisfy the RG equation
\begin{align}
 \frac{df_n(\omega_r',\mu)}{d\ln\mu} = &
- \left[\Gamma_{\rm cusp}(\alpha_s) \, \ln \frac{\mu}{\hat\omega_r'} + \gamma_+(\alpha_s)
+ \frac{\alpha_s C_F}{4\pi} \, \gamma_n^{\rm ERBL} \right]
 f_n(\omega_r',\mu)
 \cr & \qquad -
 \Gamma_{nm}(\alpha_s) \, f_m(\omega_r',\mu)\,,
 \label{fnRGE}
\end{align}
with $\gamma_n^{\rm ERBL}$ given in (\ref{gammanERBL}), and a non-diagonal contribution from
the substitution of variables in the LN kernel, given by
\begin{align}
 \Gamma_{nm}(\alpha_s) 
&=  \Gamma_{\rm cusp}(\alpha_s) \, \frac{2(2n+3)}{(n+1)(n+2)}\,
 \int\limits _0^1 du' \, u'  \bar u'\,
   \ln (u'\bar u') \, C_{n}^{(3/2)}(2u'-1) \, C_{m}^{(3/2)}(2u'-1) 
      \nonumber \\[0.5em]
      &=-\Gamma_{\rm cusp}(\alpha_s) \left( \begin{array}{cccc}
             \frac{5}{6} & \frac{3}{10} & \frac{3}{28} & \cdots \\
             \frac{7}{60} & \frac{473}{420} & \frac{7}{18} & \cdots \\
             \frac{11}{420} & \frac{11}{45} & \frac{16.847}{13.860} & \cdots \\
             \cdots & \cdots & \cdots & \cdots
            \end{array}
            \right) \,,
\end{align}
where the different lines refer to $n=0,2,4,\text{etc}$ and the columns to $m=0,2,4,\text{etc}$. 
As one can see, the off-diagonal terms are typically smaller than the diagonal ones,
and therefore, as a first approximation could be neglected.
In that case, the particular form of the leading-twist baryon LCDA in
(\ref{wouldbeasymptotic}), which translates into
\begin{align}
 \hat\rho_2(\omega_r',u',\mu) & \to  u' \, \bar u' \, f_0(\omega_r',\mu) \,,
 \label{rhospecial}
\end{align}
and 
\begin{align}
 \phi_2(\omega_1,\omega_2,\mu)
 & \to
 \frac{\omega_1 \omega_2}{\omega_1+\omega_2} \,
 \int_0^\infty \frac{d\omega_r'}{(\omega_r')^2} \,
  \sqrt{\frac{\omega_r'}{\omega_1+\omega_2}} \,
  J_3\left(2 \,\sqrt{\frac{\omega_1+\omega_2}{\omega_r'}} \right)  f_0(\omega_r',\mu)\,,
\end{align}
and
\begin{align}
 \psi_v(x_1+x_2,\mu)
 & \to \frac{1}{(x_1+x_2)^{5/2}} \, \int_0^\infty d\omega_r' \, \frac{1}{(\omega_r')^{5/2}}
 \, J_5\left(2 \,\sqrt{\frac{x_1+x_2}{\omega_r'}} \right) f_0(\omega_r',\mu) \,,
\end{align}
would be stable under evolution.
Diagonalizing the r.h.s.\ of the RG equation, truncated to a finite number of Gegenbauer coefficients,
is now also a straightforward task, which will be illustrated below.
As already discussed in \cite{Ball:2008fw} the numerical effect of the ERBL term
is in any case expected to be sub-leading, and for practical applications it should be sufficient to
treat it in an approximate way.

We finally note that the connection between the function $\tilde\phi_2(\tau_1,\tau_2)$,
appearing in the light-cone matrix elements in coordinate space,
and the spectral function $\hat\rho_2(\omega_r',u')$ in the baryonic case is given by
\begin{align}
   \hat\rho_2(\omega_r',u',\mu)
  &=\int \frac{d\tau_1}{2\pi} \, \frac{d\tau_2}{2\pi}
  \left(1- \exp\left[-\frac{i \bar u'}{\omega_r' \tau_1}\right]\right)
   \left(1- \exp\left[-\frac{i u'}{\omega_r' \tau_2}\right]\right) \tilde\phi_2(\tau_1,\tau_2,\mu) \,.
\end{align}

\subsubsection{Numerical Examples and Asymptotic Form}

\begin{figure}[t!bph]
\begin{center}
\includegraphics[width=0.48\textwidth]{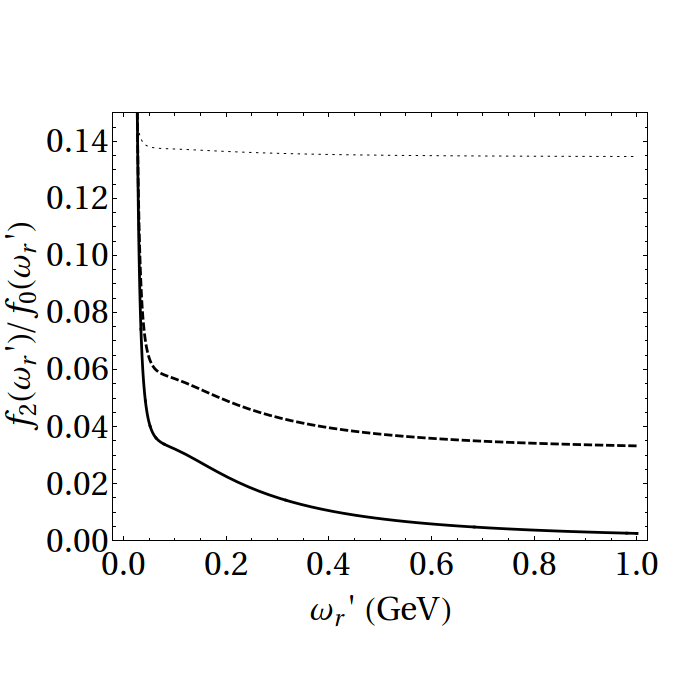}
\quad
\includegraphics[width=0.48\textwidth]{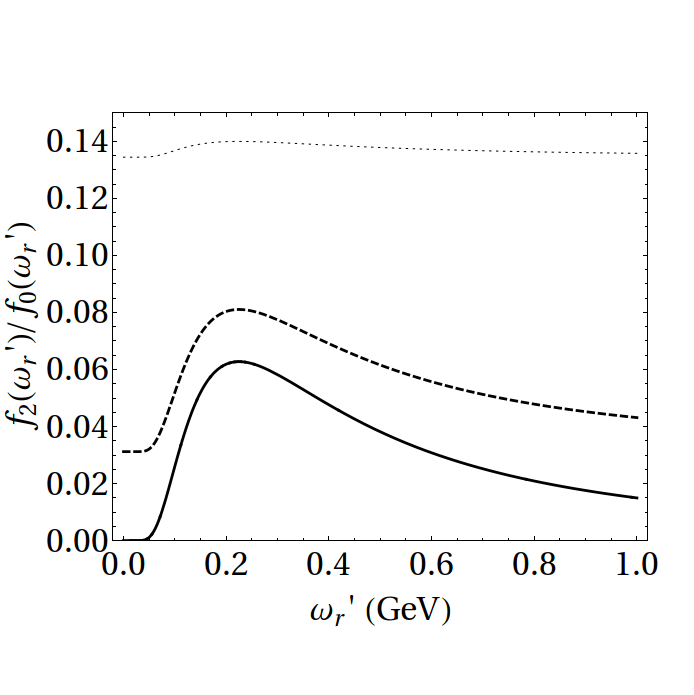}
\end{center}
\caption{\label{fig:f2f0}
Numerical examples for the evolution of the ratio of Gegenbauer coefficient
functions $f_2(\omega_r',\mu)/f_0(\omega_r',\mu)$. Left: for model~2 from \cite{Ball:2008fw}. Right: for model~2',
with $a_2=1/3$, $\epsilon_0=0.2$~GeV, $\epsilon_1=0.65$~GeV, see text.
Solid lines: input model at $\mu_0$; dashed lines: RG evolution corresponding to $g=0.3$;
dotted line: with $g=3$.}
\end{figure}

In the following, we study the
coefficient functions $f_{0,2}(\omega_r',\mu)$ and their RG behaviour, 
starting from different models
for the LCDA $\phi_2$ defined at some
input scale $\mu_0$.
For a given LCDA, making use of (\ref{In:res}) in the Appendix,
we find (using $\omega_1=u\omega$, $\omega_2=\bar u\omega$)
\begin{align}
f_n(\omega_r',\mu_0)
&= \frac{4 \,(2n+3)}{(n+1)(n+2)}
\int\limits_0^\infty \frac{d\omega}{\omega }  \sqrt{\frac{\omega}{\omega_r'}}   \,
J_{2n+3}\left(2 \sqrt{\frac{\omega}{\omega_r'}}\right)
\int\limits_0^1 du  \, C_n^{(3/2)}(2u-1) \, \phi_2(u\omega,\bar u\omega,\mu_0)
\end{align}
Notice that the Gegenbauer expansion of the original LCDA $\phi_2(u\omega,\bar u\omega)$
directly translates to the Gegenbauer expansion of the spectral function $\hat\rho_2(\omega_r',u')$.
For the models discussed above (\ref{Lambdabmodels}), this leads to
\begin{align}
\mbox{model~1:} \quad &  f_0(\omega_r',\mu_0) = \frac{1}{(\omega_r')^2} \,e^{- \omega_0/\omega_r'}
\,, \qquad  f_{n\geq 2}(\omega_r',\mu_0)=0 \,;
\cr
\mbox{model~2:} \quad &  f_0(\omega_r',\mu_0) =  \frac{1}{(\omega_r')^2} \,
e^{- \epsilon_0/\omega_r'} \,, \qquad f_{n>2}(\omega_r',\mu_0) = 0 \,,\cr
& f_2(\omega_r',\mu_0)= { a_2 \over (\omega_r')^2} \bigg \{ e^{-{1/ x_r}}
( 720 \, x_r^5 + 600 \, x_r^4 + 240 \, x_r^3 + 60 \, x_r^2 +10 \, x_r +1  )
\cr
& \phantom{ f_0(\omega_r',\mu_0) = } \qquad
 + 120 \, x_r^4 \, (1- 6 \, x_r)\bigg \} \,, \quad (x_r\equiv \omega_r'/\epsilon_1) \,;
\cr
\mbox{model~3:} \quad &  f_0(\omega_r',\mu_0) = \frac{15}{\bar \Lambda^2} \,
\sqrt{{2 \omega_r' \over \bar{\Lambda} }} \,  J_5 \left(2 \sqrt{{2 \bar{\Lambda} \over\omega_r'  }} \right) \,,
\qquad f_{n\geq 2}(\omega_r',\mu_0)=0 \,.
 \end{align}
The qualitative behaviour of model~1 and model~3 is similar to what we have discussed
for the corresponding functions in the $B$-meson case.
Notice that the contribution of the coefficient function $f_2(\omega_r')$
to the spectral function $\hat\rho_2(\omega_r',u')$ in model~2 is concentrated at very low values of $\omega_r'$, while for generic
values of $\omega_r'$ its contribution is practically negligible. In order to study
systematic deviations from 
the particular form of $\hat\rho_2(\omega_r',u')$ in (\ref{rhospecial}), it would therefore be more
convenient to define a modified version, for which we propose
\begin{align}
\mbox{model~2':} \quad &  f_0(\omega_r',\mu_0) \equiv  \frac{1}{(\omega_r')^2} \,
e^{- \epsilon_0/\omega_r'} \,,
\qquad  f_2(\omega_r',\mu_0)\equiv \frac{a_2}{6} \, { \epsilon_1^2 \over  (\omega_r')^4} \, e^{-\epsilon_1 / \omega_r'}
\,,
\end{align}
which has the same functional form as model~2 at large $\omega_r'$ (except for 
a different normalization)
such that the $1/\omega_r'$ moment of $f_2(\omega_r')$ in model~2 and
model~2' coincide. In the original momentum space, this corresponds to a LCDA
\begin{align}
 \mbox{model~2':} \quad & \phi_2(\omega_1,\omega_2, \mu_0)
 = \frac{\omega_1\omega_2 }{\epsilon_0^4}
 \, e^{-(\omega_1+\omega_2)/\epsilon_0}
 + a_2 \, \frac{\omega_1\omega_2 \left( \omega_1^2-3 \omega_1\omega_2+\omega_2^2\right)}{\epsilon_1^6}
 \, e^{-(\omega_1+\omega_2)/\epsilon_1} \,.
\end{align}
Concerning the RG behaviour, we first note
that the explicit solution for the functions $f_{0,2}(\omega_r', \mu)$
in the absence of higher Gegenbauer coefficients 
reads
\begin{align}
f_0(\omega_r',\mu) &= e^V  \left ( {\mu_0 \over \hat\omega_r'} \right )^{-g} \left\{ 
c_1 \, e^{0.877 g}
- 0.378 \, c_2 \, e^{0.040 g}  \right\} \,,
\cr
f_2(\omega_r',\mu) &= e^V  \left ( {\mu_0 \over \hat\omega_r'} \right )^{-g} \left\{ 
0.147 \, c_1 \, e^{0.877 g}
+  c_2 \, e^{0.040 g}  \right\} \,,
\end{align}
where the two integration constants are related to the initial condition of the evolution via
\begin{align}
c_1 &= 0.947 \,f_0(\omega_r',\mu_0) + 0.358 \,f_2(\omega_r',\mu_0) \,,
\cr
c_2 &=  -0.139\, f_0(\omega_r',\mu_0) + 0.947 \,f_2(\omega_r',\mu_0)  \,.
\end{align}
In the asymptotic limit, i.e.\ for large renormalization scales and large values of $g$, 
the first exponential in the curly
brackets dominates, and the ratio of the two coefficients approaches a constant, 
$f_2/f_0\simeq 0.147$. This is illustrated for model~2 and model~2' 
in Fig.~\ref{fig:f2f0}. For both models, the asymptotic value for $f_2/f_0$ 
is reached\footnote{In practice, this is
limited by the fact that the evolution of the LCDAs within HQET has to be
replaced by the standard QCD evolution above $\mu \simeq m_b$.} for $g \gsim 3$.

\begin{figure}[t!bph]
\begin{center}
\includegraphics[width=0.58\textwidth]{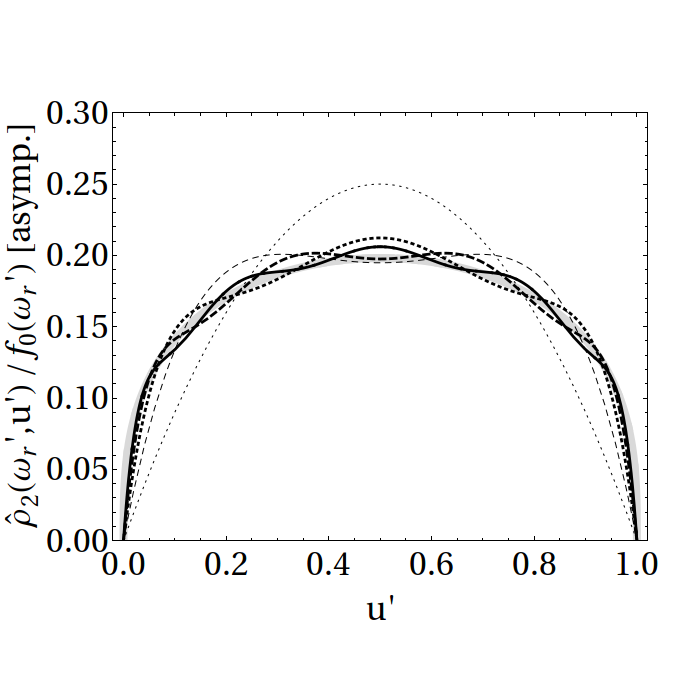}
\end{center}
\caption{\label{fig9}
Asymptotic $u'$-dependence of the spectral function
$
  \hat\rho_2(\omega_r',u',\mu)/f_0(\omega_r',\mu)
$,
for different levels of truncation in the Gegenbauer expansion: $n=0$ (thin dotted),
$n=2$ (thin dashed), $n=4$ (thick dotted), $n=6$ (thick dashed),
$n=8$ (solid). The gray band shows the approximation $\hat\rho_2(\omega_r',u',\mu)/f_0(\omega_r',\mu)
\propto (u'(1-u'))^{1/3}$.
}
\end{figure}

When higher Gegenbauer moments are included,
the asymptotic form is similarly determined by the largest eigenvalue of the 
RG equation (\ref{fnRGE}) after subtracting the LN terms.  
We then find that the ratio $f_2/f_0$ converges to about 20\%, and
that the admixtures of the higher Gegenbauer moments are less important,
with $f_4/f_0\simeq 9\%$, $f_6/f_0\simeq 5\%$, etc.
The resulting asymptotic $u'$-dependence of the spectral function, 
corresponding to the different levels of truncation in the Gegenbauer
expansion, is illustrated in Fig.~\ref{fig9}. 
The functional form
that is approached asymptotically is well approximated by 
$\hat\rho_2(\omega_r',u',\mu) \propto (u'(1-u'))^{1/3} f_0(\omega_r',\mu)$.

\clearpage

\section{Summary}

\label{summary}

We have investigated
light-cone distribution amplitudes (LCDAs) as defined in
heavy-quark effective theory for
$B$-mesons and $\Lambda_b$-baryons.
On the one hand, we have constructed easy-to-use momentum-space representations 
for the leading Fock states, which reduce to an expansion in terms of
conventional LCDAs when convoluted with a hard-scattering kernel
in the (collinear) QCD factorization approach, but also allows for
a comparison with models for transverse-momentum-dependent
wave functions. In the simplest case, our construction automatically implements
so-called Wandzura-Wilczek relations, which connect different
LCDAs in the limit where higher Fock-state contributions to
the equations of motion are neglected. We have also illustrated
how corrections to the Wandzura-Wilczek approximation can be taken into account
consistently within our approach. For the baryonic case in
particular, our ansatz leads to a significant reduction of independent
hadronic functions which appear at sub-leading order in the collinear expansion.
The sub-leading functions are needed, for instance, in SCET sum rules analyses
of exclusive $\Lambda_b$ transitions for cases where the standard
QCD factorization approach would lead to endpoint-sensitive (formally
ill-defined) convolution integrals (see e.g.\ the discussion in
\cite{Feldmann:2011xf}).

Furthermore, we have found a new representation of LCDAs 
in terms of dual spectral functions, which are 
the eigenfunctions of the Lange-Neubert renormalization kernel.
The connection between
the LCDAs and their dual representations is via convolution integrals with
Bessel functions.
In the dual space, the solutions to the
renormalization-group equations are extremely simple.
In the mesonic case, 
they are local in the dual momentum variable
$\omega'$ 
of the light quark. 
We have demonstrated the simplifications that arise when 
re-formulating the factorization theorem for radiative leptonic $B \to \gamma\ell\nu$
decays in terms of the new spectral function and an associated new 
hard-collinear function, which evolve both by a multiplicative factor.
Again, the baryonic case is more complicated, because the Lange-Neubert
kernel and the Efremov-Radyushkin--Brodsky-Lepage kernel cannot be
separated in the usual momentum space. In the dual space, however, 
a separation of the reduced dual momentum 
$\omega_r'=\omega_1'\omega_2'/(\omega_1'+\omega_2')$
and the dual momentum fraction $u' = \omega_1'/(\omega_1'+\omega_2')$
is possible, which opens the way for a systematic solution of the 
baryonic renormalization-group equation
in terms of an expansion in Gegenbauer polynomials (as known from the
pion LCDAs).

In summary, our results should be helpful in calculations of
exclusive decay amplitudes for hadrons containing a heavy quark
in the framework of QCD factorization, soft-collinear effective theory
or light-cone sum rules. Specifically for applications involving
heavy baryons, we expect
more transparent and efficient estimates of theoretical hadronic uncertainties,
 which
are needed, for instance, to constrain physics beyond the Standard Model
from rare decays like $\Lambda_b \to \Lambda \mu^+\mu^-$ which are
currently studied at hadron colliders.

\clearpage

\section*{Acknowledgements}
We would like to thank Martin Beneke and Bj\"orn Lange for a critical reading of the
manuscript and helpful comments.
GB gratefully acknowledges the support of a University Research Fellowship
by the Royal Society.
TF would like to thank Danny van Dyk for helpful comments and questions.
TF further acknowledges support by the Deutsche Forschungsgemeinschaft (DFG)
within the Research Unit {\sc FOR 1873} ({\it Quark Flavour Physics and Effective Field Theories}).
YMW would like to thank Grigory Kirilin for discussions on integrals
with Bessel functions. YMW is supported
by the  {\it DFG-Sonder\-forschungs\-bereich/Trans\-regio 9 ``Computer\-gest\"{u}tzte Theoretische Teilchen\-physik"}.
MWYY is supported by a Durham University Doctoral Fellowship.
He would also like to thank
the University of Siegen for financial support during his visits in 2012 and 2013.


\appendix
\section{$B$-Meson Projectors in arbitrary Frame}

In certain applications, one would need the momentum-space projectors in an arbitrary frame,
where $n_- \cdot n_+ = 2$, but the heavy-quark velocity reads
$$v^\mu = (n_-v) \, \frac{n_+^\mu}{2} + (n_+v) \, \frac{n_-^\mu}{2} + v_\perp^\mu\,,
\qquad v_\perp^2 = 1 - (n_-v)(n_+v) \,. $$

\subsection{2-Particle Projector}

\label{generalframe}

Taking $(n_-z) \ll z_\perp \ll (n_+z)$ and expanding the non-local matrix element for the $B$-meson
2-particle LCDAs,
we then obtain
\begin{align}
& 2 \, \tilde \Phi_B^+(t,z^2) +
  \frac{\tilde \Phi_B^-(t,z^2) - \tilde \Phi_B^+(t,z^2)}{t} \, \slashed z
  \cr
&\qquad= 2 \, \tilde \phi_B^+(\tau) +
  \left( \tilde \phi_B^-(\tau) - \tilde \phi_B^+(\tau) \right)  \frac{\slashed n_-}{(n_-v)} +
  \frac{\tilde \phi_B^-(\tau) - \tilde \phi_B^+(\tau)}{\tau}
  \, \slashed z_\perp
\cr
&\qquad \quad  + \frac{\partial}{\partial\tau} \left(
   2  \tilde \phi_B^+(\tau) + \left( \tilde \phi_B^-(\tau) - \tilde \phi_B^+(\tau) \right) \frac{\slashed n_-}{(n_-v)}
   \right) (v_\perp \cdot z_\perp)
+ {\cal O}(z_\perp^2, n_-z) \,,
\label{lightBarb} 
\end{align}
where now $\tau = \frac{(n_-v) \, (n_+ z)}{2}$ can be interpreted as
the Fourier-conjugated variable to the momentum component  of the light anti-quark,
$$
\omega=\frac{(n_- k)}{(n_-v)} \,, \qquad
 \omega \, \tau  = \frac{(n_-k)(n_+z)}{2} \,.
$$
The light-cone expansion in (\ref{lightBarb}) corresponds to the momentum-space projector
\begin{align}
 & 2\phi_B^+(\omega) 
 + \Big( \phi_B^-(\omega) - \phi_B^+(\omega) \Big) \frac{\slashed n_-}{(n_-v)}
  -\int_0^\omega d\eta \,\Big(\phi_B^-(\eta) - \phi_B^+(\eta)\Big) \,\gamma^\mu_{\perp}\, \frac{\partial}{\partial k_\perp^\mu }
  \cr
  & \qquad {} +  \omega \bigg( 
  2\phi_B^+(\omega) 
 + \Big( \phi_B^-(\omega) - \phi_B^+(\omega)   \Big)
  \frac{\slashed n_-}{(n_-v)}
   \bigg) \left(v_\perp \cdot \frac{\partial}{\partial k_\perp}\right)
   \cr &
\label{boostinv}
\end{align}
to be used in factorization theorems
with hard-scattering kernels where the $k_\perp^2$ and $(n_+k)$-dependence can be neglected.
Notice that the so-defined projector is manifestly invariant under Lorentz boosts, $n_\pm \to \gamma^{\pm 1} n_\pm$.

\subsection{3-Particle Projector}

\label{3general}

Following the same procedure as for the 2-particle momentum-space projector,
we obtain the leading contribution to the 3-particle  projector in a general frame as 
\begin{align}
{\cal M}_B^{(3)}{}^\mu(v,\omega,\xi) &=- {i \over \xi}  {\tilde{f}_B m_B \over 2}\, \Bigg [ {1 + \slashed v \over 2 }
\Bigg\{ 
\cr & \quad
\left(
\frac{ (n_{+}  v)  \,  n_{-}^{\mu}  \slashed n_{-} - n_-^\mu
+ (v_{\perp}^{\mu} \slashed n_{-}+ n_{-}^{\mu} \slashed v_{\perp})}{(n_-v)}- \gamma_{\perp}^{\mu}     \right)
\big(\Psi_A(\omega,\xi) - \Psi_V(\omega,\xi)\big)
\cr
& \quad
+ \left( 
\frac{ n_{-}^{\mu}  \slashed n_{-} + n_{-}^{\mu} \slashed n_{-} \slashed v_{\perp}}{(n_-v)}
-
n_{-}^{\mu}  + \gamma_{\perp}^{\mu}\slashed n_{-}\right)
 \frac{\Psi_V(\omega,\xi)}{(n_{-} v)}
\cr
& \quad - \frac{ n_{-}^{\mu}}{(n_-v)}\,X_A(\omega,\xi) 
+ \frac{ n_{-}^{\mu}  \slashed n_{-}}{(n_-v)^2} \,Y_A(\omega,\xi) \Bigg\} \gamma_5 \Bigg ]\,,
\end{align}
which is again manifestly invariant under longitudinal Lorentz boosts.

\section{The ERBL Term for the dual LCDA of the $\Lambda_b$--Baryon}

\label{app:ERBL}

The ERBL contribution to the RG equation for the LCDAs $\phi_2(\omega_1,\omega_2)$ reads \cite{Ball:2008fw}
\begin{align}
 \frac{d}{d\ln\mu} \, \phi_2(u\omega,\bar u\omega,\mu) \Big|_{\rm ERBL}
 &=  \frac{\alpha_s C_F}{2\pi} \, \int_0^1 dv \, V^{\rm ERBL}(u,v) \, \phi_2(v\omega,\bar v\omega,\mu) \,,
\end{align}
with
\begin{align}
\omega=\omega_1+\omega_2 \,, \quad
u = \frac{\omega_1}{\omega_1+\omega_2} \,, \quad \bar u= 1-u= \frac{\omega_2}{\omega_1+\omega_2} \,.
\end{align}
The one-loop expression for the ERBL kernel is given by
\begin{align}
 V^{\rm ERBL}(u,v) &= \left[\frac{1-u}{1-v} \left( 1 + \frac{1}{u-v} \right) \theta(u-v) + \frac{u}{v}
 \left( 1 + \frac{1}{v-u} \right) \theta(v-u) \right]_+
 \cr
 &= - u(1-u) \sum_{n=0}^\infty \frac{2(2n+3)}{(n+1)(n+2)} \, \gamma_n^{\rm ERBL} \,
   C_n^{3/2}(2u-1) \, C_n^{3/2}(2v-1) \,,
\end{align}
where in the last line we have quoted the expansion in terms of Gegenbauer polynomials
\cite{Braun:2003rp}, with the eigenvalues of the corresponding  anomalous-dimension matrix
given by
\begin{align}
 \gamma_n^{\rm ERBL} & = 1- \frac{2}{(n+1)(n+2)} + 4 \,\sum_{m=2}^{n+1} \frac{1}{m} \,.
 \label{gammanERBL}
\end{align}
The transformation to the spectral function $\rho_2$ reads
\begin{align}
 & \frac{d}{d\ln\mu} \, \rho_2(\omega_1',\omega_2',\mu) \Big|_{\rm ERBL} \cr
 & \quad = \frac{d}{d\ln\mu}  \int\limits_0^\infty \frac{d\omega_1}{\omega_1} \int\limits_0^\infty \frac{d\omega_2}{\omega_2}
 \, \sqrt{\frac{\omega_1 \omega_2}{\omega_1'\omega_2'}} \, J_1\left( 2 \sqrt{\frac{\omega_1}{\omega_1'}} \right)
 J_1\left( 2 \sqrt{\frac{\omega_2}{\omega_2'}} \right) \phi_2(\omega_1,\omega_2,\mu)\Big|_{\rm ERBL}
 \cr
 &\quad = \frac{d}{d\ln\mu}  \int\limits_0^\infty \frac{d\omega}{\omega} \int\limits_0^1 \frac{du}{u\bar u}
 \, \sqrt{\frac{u \bar u \omega^2}{\omega_1'\omega_2'}} \, J_1\left( 2 \sqrt{\frac{u \omega}{\omega_1'}} \right)
 J_1\left( 2 \sqrt{\frac{\bar u\omega}{\omega_2'}} \right) \phi_2(u\omega,\bar u\omega,\mu)\Big|_{\rm ERBL}
 \cr & \quad = \frac{\alpha_s C_F}{2\pi} \, \int\limits_0^\infty \frac{d\omega}{\omega} \int\limits_0^1 \frac{du}{u\bar u}
 \, \sqrt{\frac{u \bar u \omega^2}{\omega_1'\omega_2'}} \, J_1\left( 2 \sqrt{\frac{u \omega}{\omega_1'}} \right)
 J_1\left( 2 \sqrt{\frac{\bar u\omega}{\omega_2'}} \right)
 \cr & \qquad \ \times  \int_0^1 dv \, V^{\rm ERBL}(u,v) \, \phi_2(v\omega,\bar v\omega,\mu)
 \cr
 & \quad = \frac{\alpha_s C_F}{2\pi} \, \int\limits_0^\infty \frac{d\omega}{\omega} \int\limits_0^1 \frac{du}{u\bar u}
 \, \sqrt{\frac{u \bar u \omega^2}{\omega_1'\omega_2'}} \, J_1\left( 2 \sqrt{\frac{u \omega}{\omega_1'}} \right)
 J_1\left( 2 \sqrt{\frac{\bar u\omega}{\omega_2'}} \right) \int_0^1 dv \, V^{\rm ERBL}(u,v)
 \cr & \qquad \ \times \int\limits_0^\infty \frac{d\eta_1'}{\eta_1'}
 \int\limits_0^\infty \frac{d\eta_2'}{\eta_2'} \,\sqrt{\frac{v \bar v \omega^2}{\eta_1'\eta_2'}} \,
  J_1\left( 2 \sqrt{\frac{v\omega}{\eta_1'}} \right) J_1\left( 2 \sqrt{\frac{\bar v\omega}{\eta_2'}} \right)
  \rho_2(\eta_1',\eta_2',\mu) \,.
\end{align}
In terms of the integrals $I_n$ defined in (\ref{In}) below,
using the Gegenbauer expansion of the ERBL kernel,
we can write
\begin{align}
 & \frac{d}{d\ln\mu} \, \rho_2(\omega_1',\omega_2',\mu) \Big|_{\rm ERBL}
\cr
 & \quad = -\frac{\alpha_s C_F}{2\pi} \,
 \sum_{n=0}^\infty \frac{2(2n+3)}{(n+1)(n+2)} \, \gamma_n^{\rm ERBL}
 \int\limits_0^\infty \frac{d\omega}{\omega} \int\limits_0^\infty \frac{d\eta_1'}{\eta_1'}
 \int\limits_0^\infty \frac{d\eta_2'}{\eta_2'}
 \,
 \cr & \quad \times \, \sqrt{\frac{\omega^2}{\omega_1'\omega_2'}} \,
 \sqrt{\frac{\omega^2}{\eta_1'\eta_2'}}\,
 I_n\left(2 \sqrt{\frac{\omega}{\omega_1'}},2 \sqrt{\frac{\omega}{\omega_2'}}\right)
  \,
 I_n\left(  2 \sqrt{\frac{\omega}{\eta_1'}}, 2 \sqrt{\frac{\omega}{\eta_2'}} \right)
  \rho_2(\eta_1',\eta_2',\mu) \,.
\end{align}
As the integrals $I_n$ themselves are proportional to Bessel functions and
have a homogeneous scaling with the variable $\omega$, we can use the completeness
relation (\ref{complete}) to perform the $\omega$ integration explicitly for
each individual order in the Gegenbauer expansion.
It is furthermore convenient to introduce new variables
\begin{align} \omega_r' \equiv \frac{\omega_1' \omega_2'}{\omega_1'+\omega_2'} & \qquad \mbox{and}
\quad u'=1-\bar u' = \frac{\omega_1'}{\omega_1'+\omega_2'} \,,
\cr
\eta_r' \equiv \frac{\eta_1' \eta_2'}{\eta_1'+\eta_2'} & \qquad \mbox{and}
\quad v'=1-\bar v' = \frac{\eta_1'}{\eta_1'+\eta_2'} \,,
\end{align}
and to denote the spectral function in terms of the new variables according to
\begin{align}
  \rho_2(\omega_1',\omega_2')\equiv \hat \rho_2(\omega_r',u') \,.
\end{align}
We then obtain
\begin{align}
 & \frac{d}{d\ln\mu} \, \hat\rho_2(\omega_r',u',\mu) \Big|_{\rm ERBL}
\cr
&\qquad= - \frac{\alpha_s C_F}{2\pi} \,
 \sum_{n=0}^\infty \frac{2(2n+3)}{(n+1)(n+2)} \,
  \gamma_n^{\rm ERBL} \,
u' \bar u' \, C_n^{(3/2)}(2u'-1) 
\cr
& \qquad\qquad \times
\int_0^1 dv' \,  C_n^{(3/2)}(2v'-1) \,\int_0^\infty d\eta_r' \, 
 \delta(\omega_r'-\eta_r') \, \hat\rho_2(\eta_r',v',\mu)
 \cr
 &\qquad= \frac{\alpha_s C_F}{2\pi} \, \int_0^1 dv' \, V^{\rm ERBL}(u',v') \, \hat\rho_2(\omega_r',v',\mu) \,.
 \end{align}

\section{Some Relations with Bessel Functions}

The completeness relation for Bessel functions,
\begin{align*}
  \int_0^\infty dz \, z \, J_n(az) \, J_n(bz) &= \frac{1}{a} \,
\delta(a-b) \,,
\end{align*}
can be written as
\begin{align}
& \int_0^\infty \frac{d\omega'}{\omega'} \, \frac{1}{\omega'} \,
J_{n}\left( 2 \, \sqrt{\frac{a}{\omega'} }\right)
   J_{n}\left( 2 \, \sqrt{\frac{b}{\omega'}}\right)
   \cr
   = &\int_0^\infty d\omega \, J_{n}\left( 2 \, \sqrt{a \omega }\right)
   J_{n}\left( 2 \, \sqrt{ b  \omega}\right)
=\delta(a-b)\,,
   \label{complete}
\end{align}
which has been frequently used in the text.

\noindent 
We further define integrals with Bessel functions and
Gegenbauer polynomials,
\begin{align}
  I_n(\alpha,\beta) &\equiv \int_0^1 dv \, \sqrt{v(1-v)} \, J_1(\alpha
\sqrt{v}) \, J_1(\beta\sqrt{1-v}) \,
     C_n^{3/2}(2v-1) \,.
     \label{In}
\end{align}
For the first few (even) values of $n$, we obtain
\begin{align}
  I_0(\alpha,\beta) &= 2 \, \alpha\beta
\frac{1}{(\alpha^2+\beta^2)^{3/2}} \, J_3(\sqrt{\alpha^2+\beta^2}) \,,
  \\
  I_2(\alpha,\beta) &=
  12 \, \alpha \beta \, \frac{\alpha^4-3 \, \alpha^2
\beta^2+\beta^4}{(\alpha^2+\beta^2)^{7/2}} \,
  J_7(\sqrt{\alpha^2+\beta^2}) \,,
  \\
  I_4(\alpha,\beta) &=
  30 \, \alpha \beta \,
  \frac{\alpha^8-10\, \alpha^6\beta^2+ 20\, \alpha^4\beta^4 - 10\,
\alpha^2\beta^6 + \beta^8}{(\alpha^2+\beta^2)^{11/2}} \,
  J_{11}(\sqrt{\alpha^2+\beta^2})
  \,.
\end{align}
The general formula can be constructed by introducing the variables
\begin{align}
  \alpha = 2 \, \sqrt{\frac{x}{1-u}} \,, \quad \beta = 2 \,
\sqrt{\frac{x}{u}} \,,
  \quad \Leftrightarrow \quad
  x = \frac{\alpha^2\beta^2}{4 \, (\alpha^2+\beta^2)} \,, \qquad u =
\frac{\alpha^2}{\alpha^2 + \beta^2} \,,
\end{align}
for which we obtain the compact expression
\begin{align}
I_n(\alpha,\beta) &= \frac{u(1-u)}{\sqrt x} \, C_n^{(3/2)}(2u-1) \, J_{2n+3}\left( 2
\sqrt{\frac{x}{u(1-u)}} \right) \,.
\label{In:res}
\end{align}

\noindent We also often used the relation
\begin{align}
  J_n(x) &=  \frac{x}{2n} \left( J_{n-1}(x) + J_{n+1}(x) \right) \qquad
(n \geq 1) \,.
\end{align}



\begin{thebibliography}{99}

\bibitem{Efremov:1979qk}
  A.~V.~Efremov and A.~V.~Radyushkin,
  ``Factorization and Asymptotical Behavior of Pion Form-Factor in QCD,''
  Phys.\ Lett.\ B {\bf 94} (1980) 245.

\bibitem{Lepage:1979zb}
  G.~P.~Lepage and S.~J.~Brodsky,
  ``Exclusive Processes in Quantum Chromodynamics: Evolution Equations for Hadronic Wave Functions and the Form-Factors of Mesons,''
  Phys.\ Lett.\ B {\bf 87} (1979) 359;
  ``Exclusive Processes in Perturbative Quantum Chromodynamics,''
  Phys.\ Rev.\ D {\bf 22} (1980) 2157.

\bibitem{Duncan:1979hi}
  A.~Duncan and A.~H.~Mueller,
  ``Asymptotic Behavior of Composite Particle Form-Factors and the Renormalization Group,''
  Phys.\ Rev.\ D {\bf 21} (1980) 1636.

\bibitem{Chernyak:1983ej}
  V.~L.~Chernyak and A.~R.~Zhitnitsky,
  ``Asymptotic Behavior of Exclusive Processes in QCD,''
  Phys.\ Rept.\  {\bf 112} (1984) 173.

\bibitem{Korchemsky:1999qb}
  G.~P.~Korchemsky, D.~Pirjol and T.~-M.~Yan,
  ``Radiative leptonic decays of $B$ mesons in QCD,''
  Phys.\ Rev.\ D {\bf 61} (2000) 114510
  [hep-ph/9911427].

\bibitem{DescotesGenon:2002mw}
  S.~Descotes-Genon and C.~T.~Sachrajda,
  ``Factorization, the light cone distribution amplitude of the $B$-meson and the radiative decay
  $B \to \gamma \ell \nu_\ell$,''
  Nucl.\ Phys.\ B {\bf 650} (2003) 356
  [hep-ph/0209216].

\bibitem{Lunghi:2002ju}
  E.~Lunghi, D.~Pirjol and D.~Wyler,
  ``Factorization in leptonic radiative $b \to \gamma e \nu$ decays,''
  Nucl.\ Phys.\ B {\bf 649} (2003) 349
  [hep-ph/0210091].

\bibitem{Bosch:2003fc}
  S.~W.~Bosch, R.~J.~Hill, B.~O.~Lange and M.~Neubert,
  ``Factorization and Sudakov resummation in leptonic radiative $B$ decay,''
  Phys.\ Rev.\ D {\bf 67} (2003) 094014
  [hep-ph/0301123].

\bibitem{Beneke:2011nf}
  M.~Beneke and J.~Rohrwild,
  ``$B$-meson distribution amplitude from $B \to \gamma \ell \nu$,''
  Eur.\ Phys.\ J.\ C {\bf 71} (2011) 1818
  [arXiv:1110.3228 [hep-ph]].

\bibitem{Braun:2012kp}
  V.~M.~Braun and A.~Khodjamirian,
  ``Soft contribution to $B\to \gamma \ell \nu_\ell$ and the $B$-meson distribution amplitude,''
  Phys.\ Lett.\ B {\bf 718} (2013) 1014
  [arXiv:1210.4453 [hep-ph]].

\bibitem{Beneke:1999br}
  ``QCD factorization for $B \to \pi \pi$ decays: Strong phases and CP violation in the heavy quark limit,''
  Phys.\ Rev.\ Lett.\  {\bf 83} (1999) 1914
  [hep-ph/9905312];
  M.~Beneke, G.~Buchalla, M.~Neubert and C.~T.~Sachrajda,
  ``QCD factorization in $B \to \pi K, \pi \pi$ decays and extraction of Wolfenstein parameters,''
  Nucl.\ Phys.\ B {\bf 606} (2001) 245
  [hep-ph/0104110].



\bibitem{Beneke:2000wa}
  M.~Beneke, Th.~Feldmann,
  ``Symmetry breaking corrections to heavy-to-light $B$-meson form factors at large recoil,''
  Nucl.\ Phys.\  {\bf B592}, 3-34 (2001)
  [hep-ph/0008255].

\bibitem{Bosch:2001gv}
  S.~W.~Bosch and G.~Buchalla,
  ``The Radiative decays $B \to V \gamma$ at next-to-leading order in QCD,''
  Nucl.\ Phys.\ B {\bf 621} (2002) 459
  [hep-ph/0106081].

\bibitem{Ali:2001ez}
  A.~Ali and A.~Y.~Parkhomenko,
  ``Branching ratios for $B \to K^* \gamma$ and $B \to \rho \gamma$
   decays in next-to-leading order in the large energy effective theory,''
  Eur.\ Phys.\ J.\ C {\bf 23} (2002) 89
  [hep-ph/0105302];
  A.~Ali, B.~D.~Pecjak and C.~Greub,
  ``$B \to V \gamma$ Decays at NNLO in SCET,''
  Eur.\ Phys.\ J.\ C {\bf 55} (2008) 577
  [arXiv:0709.4422 [hep-ph]].



\bibitem{Beneke:2001at}
  M.~Beneke, Th.~Feldmann and D.~Seidel,
  ``Systematic approach to exclusive $B \to V \ell^+ \ell^-$, $V \gamma$ decays,''
  Nucl.\ Phys.\ B {\bf 612} (2001) 25
  [hep-ph/0106067];
  ``Exclusive radiative and electroweak $b \to d$ and $b \to s$ penguin decays at NLO,''
  Eur.\ Phys.\ J.\ C {\bf 41} (2005) 173
  [hep-ph/0412400].

\bibitem{Kagan:2001zk}
  A.~L.~Kagan and M.~Neubert,
  ``Isospin breaking in $B \to K^* \gamma$ decays,''
  Phys.\ Lett.\ B {\bf 539} (2002) 227
  [hep-ph/0110078].


\bibitem{Bauer:2000yr}
  C.~W.~Bauer, S.~Fleming, D.~Pirjol and I.~W.~Stewart,
  ``An Effective field theory for collinear and soft gluons: Heavy to light decays,''
  Phys.\ Rev.\ D {\bf 63} (2001) 114020
  [hep-ph/0011336];
  C.~W.~Bauer, D.~Pirjol and I.~W.~Stewart,
  ``Soft collinear factorization in effective field theory,''
  Phys.\ Rev.\ D {\bf 65} (2002) 054022
  [hep-ph/0109045].

\bibitem{Beneke:2002ph}
  M.~Beneke, A.~P.~Chapovsky, M.~Diehl, Th.~Feldmann,
  ``Soft collinear effective theory and heavy to light currents beyond leading power,''
  Nucl.\ Phys.\  {\bf B643 } (2002)  431-476
  [hep-ph/0206152];
  M.~Beneke and Th.~Feldmann,
  ``Multipole expanded soft collinear effective theory with non-Abelian gauge symmetry,''
  Phys.\ Lett.\ B {\bf 553} (2003) 267
  [hep-ph/0211358].


\bibitem{DeFazio:2005dx}
  F.~De Fazio, Th.~Feldmann, T.~Hurth,
  ``Light-cone sum rules in soft-collinear effective theory,''
  Nucl.\ Phys.\  {\bf B733 } (2006)  1-30
  [hep-ph/0504088];
  ``SCET sum rules for $B \to P$ and $B \to V$ transition form factors,''
  JHEP {\bf 0802 } (2008)  031
  [arXiv:0711.3999 [hep-ph]].

\bibitem{Khodjamirian:2005ea}
  A.~Khodjamirian, T.~Mannel and N.~Offen,
  ``$B$-meson distribution amplitude from the $B \to \pi$ form-factor,''
  Phys.\ Lett.\ B {\bf 620} (2005) 52
  [hep-ph/0504091];
  ``Form-factors from light-cone sum rules with $B$-meson distribution amplitudes,''
  Phys.\ Rev.\  {\bf D75 } (2007)  054013.
  [hep-ph/0611193].



\bibitem{Keum:2000wi}
  Y.~Y.~Keum, H.~-N.~Li and A.~I.~Sanda,
  ``Penguin enhancement and $B \to K \pi$ decays in perturbative QCD,''
  Phys.\ Rev.\ D {\bf 63} (2001) 054008
  [hep-ph/0004173].


\bibitem{Lu:2000em}
  C.~-D.~Lu, K.~Ukai and M.~-Z.~Yang,
  ``Branching ratio and CP violation of $B \to \pi \pi$ decays in perturbative QCD approach,''
  Phys.\ Rev.\ D {\bf 63} (2001) 074009
  [hep-ph/0004213].

\bibitem{He:2006ud}
  X.~-G.~He, T.~Li, X.~-Q.~Li and Y.~-M.~Wang,
  ``PQCD calculation for $\Lambda_b \to \Lambda \gamma$ in the standard model,''
  Phys.\ Rev.\ D {\bf 74} (2006) 034026
  [hep-ph/0606025];
  C.~-D.~Lu, Y.~-M.~Wang, H.~Zou, A.~Ali and G.~Kramer,
  ``Anatomy of the pQCD Approach to the Baryonic Decays $\Lambda_b\to p \pi, p K$,''
  Phys.\ Rev.\ D {\bf 80} (2009) 034011
  [arXiv:0906.1479 [hep-ph]].



\bibitem{Grozin:1996pq}
  A.~G.~Grozin and M.~Neubert,
  ``Asymptotics of heavy-meson form factors,''
  Phys.\ Rev.\  D {\bf 55} (1997) 272
  [arXiv:hep-ph/9607366];
  A.~G.~Grozin,
  ``$B$-meson distribution amplitudes,''
  Int.\ J.\ Mod.\ Phys.\  A {\bf 20} (2005) 7451
  [arXiv:hep-ph/0506226].



\bibitem{Kawamura:2001jm}
  H.~Kawamura, J.~Kodaira, C.~-F.~Qiao, K.~Tanaka,
  ``$B$ meson light cone distribution amplitudes in the heavy quark limit,''
  Phys.\ Lett.\ B {\bf 523} (2001) 111
   [Erratum-ibid.\ B {\bf 536} (2002) 344]
  [hep-ph/0109181].



\bibitem{Ball:2008fw}
  P.~Ball, V.~M.~Braun and E.~Gardi,
  ``Distribution Amplitudes of the $\Lambda_b$ Baryon in QCD,''
  Phys.\ Lett.\ B {\bf 665} (2008) 197
  [arXiv:0804.2424 [hep-ph]].


\bibitem{Feldmann:2011xf}
  Th.~Feldmann and M.~W.~Y.~Yip,
  ``Form Factors for $\Lambda_b \to \Lambda$ Transitions in SCET,''
  Phys.\ Rev.\ D {\bf 85} (2012) 014035
   [Erratum-ibid.\ D {\bf 86} (2012) 079901]
  [arXiv:1111.1844 [hep-ph]];
  M.~W.~Y.~Yip, ``Rare Decays of Heavy Baryons
using Soft Collinear Effective Theory'', PhD Thesis,
 Durham University, July 2013.


\bibitem{Ali:2012pn}
  A.~Ali, C.~Hambrock, A.~Y.~.Parkhomenko and W.~Wang,
  ``Light-Cone Distribution Amplitudes of the Ground State Bottom Baryons in HQET,''
  Eur.\ Phys.\ J.\ C {\bf 73} (2013) 2302
  [arXiv:1212.3280 [hep-ph]].



\bibitem{Lange:2003ff}
  B.~O.~Lange, M.~Neubert,
  ``Renormalization group evolution of the $B$-meson light cone distribution amplitude,''
  Phys.\ Rev.\ Lett.\  {\bf 91} (2003) 102001
  [hep-ph/0303082].
\bibitem{Braun:2003wx}
  V.~M.~Braun, D.~Y.~Ivanov and G.~P.~Korchemsky,
  ``The B-Meson Distribution Amplitude in QCD,''
  Phys.\ Rev.\  D {\bf 69} (2004) 034014
  [arXiv:hep-ph/0309330].

\bibitem{Lange:2004yh}
  B.~O.~Lange,
  ``Soft-collinear factorization and Sudakov resummation of heavy meson decay amplitudes with effective field theories,''
  hep-ph/0409277.


\bibitem{Lee:2005gza}
  S.~J.~Lee, M.~Neubert,
  ``Model-independent properties of the $B$-meson distribution amplitude,''
  Phys.\ Rev.\ D {\bf 72} (2005) 094028
  [hep-ph/0509350].

\bibitem{Bell:2008er}
  G.~Bell, Th.~Feldmann,
  ``Modelling light-cone distribution amplitudes from non-relativistic bound states,''
  JHEP {\bf 0804} (2008) 061
  [arXiv:0802.2221 [hep-ph]].


\bibitem{DescotesGenon:2009hk}
  S.~Descotes-Genon, and N.~Offen,
  ``Three-particle contributions to the renormalisation of $B$-meson light-cone distribution amplitudes,''
  JHEP {\bf 0905} (2009) 091
  [arXiv:0903.0790 [hep-ph]];
 ``Renormalization of $B$-meson distribution amplitudes,''
  PoS EFT {\bf 09} (2009) 004
  [arXiv:0904.4687 [hep-ph]].
  
\bibitem{Knodlseder:2011gc}
  M.~Knodlseder and N.~Offen,
  ``Renormalisation of heavy-light light operators,''
  JHEP {\bf 1110} (2011) 069
  [arXiv:1105.4569 [hep-ph]].

\bibitem{Pilipp:2007sb}
  V.~Pilipp,
  ``Matching of $\lambda_B$ onto HQET,''
  hep-ph/0703180.

\bibitem{Li:2012md}
  H.~-N.~Li, Y.~-L.~Shen, Y.~-M.~Wang,
  ``Resummation of rapidity logarithms in $B$ meson wave functions,''
  JHEP {\bf 1302} (2013) 008
  [arXiv:1210.2978 [hep-ph]].

\bibitem{Kawamura:2010tj}
  H.~Kawamura and K.~Tanaka,
  ``Evolution equation for the $B$-meson distribution amplitude in the heavy-quark effective theory in coordinate space,''
  Phys.\ Rev.\ D {\bf 81} (2010) 114009
  [arXiv:1002.1177 [hep-ph]].
  
\bibitem{Kawamura:2008vq}
  H.~Kawamura and K.~Tanaka,
  ``Operator product expansion for $B$-meson distribution amplitude and dimension-5 HQET operators,''
  Phys.\ Lett.\ B {\bf 673} (2009) 201
  [arXiv:0810.5628 [hep-ph]].
  
\bibitem{Collins:2003fm}
 J.~C.~Collins,
 ``What exactly is a parton density?,''
 Acta Phys.\ Polon.\ B {\bf 34} (2003) 3103
 [hep-ph/0304122].

\bibitem{Geyer:2005fb}
  B.~Geyer and O.~Witzel,
  ``$B$-meson distribution amplitudes of geometric twist vs. dynamical twist,''
  Phys.\ Rev.\ D {\bf 72} (2005) 034023
  [hep-ph/0502239].
      
\bibitem{Huang:2005kk}
  T.~Huang, C.~-F.~Qiao, and X.~-G.~Wu,
  ``$B$-meson wavefunction with 3-particle Fock states' contributions,''
  Phys.\ Rev.\ D {\bf 73} (2006) 074004
  [hep-ph/0507270].

\bibitem{Wang:2009hra}
  Y.~-M.~Wang, Y.~-L.~Shen and C.~-D.~L\"u,
  ``$\Lambda_b \to p, \Lambda$ transition form factors from QCD light-cone sum rules,''
  Phys.\ Rev.\ D {\bf 80} (2009) 074012
  [arXiv:0907.4008 [hep-ph]].

\bibitem{Wang:2011uv}
  W.~Wang,
  ``Factorization of Heavy-to-Light Baryonic Transitions in SCET,''
  Phys.\ Lett.\ B {\bf 708} (2012) 119
  [arXiv:1112.0237 [hep-ph]].
  
\bibitem{Braun:2003rp}
  V.~M.~Braun, G.~P.~Korchemsky and D.~M\"uller,
  ``The Uses of conformal symmetry in QCD,''
  Prog.\ Part.\ Nucl.\ Phys.\  {\bf 51} (2003) 311
  [hep-ph/0306057].
  
 


\end{thebibliography}
\end{document}